\documentclass[pra,twocolumn]{revtex4}

\usepackage{epsfig,amsmath}
\usepackage{subfigure}
\usepackage{graphicx}
\usepackage{dcolumn}
\usepackage{stmaryrd}
\usepackage{mathrsfs}
\usepackage{pifont}
\usepackage{amsthm}
\usepackage{amssymb}
\usepackage{bm}
\usepackage{latexsym}
\usepackage{hyperref}
\usepackage{color}

\usepackage{algorithmic}
\usepackage{algorithm}

\newcommand{\beq}{\begin{equation}}
\newcommand{\eeq}{\end{equation}}
\newcommand{\beqa}{\begin{eqnarray}}
\newcommand{\eeqa}{\end{eqnarray}}

\begin{document}

\title{{Enhanced quantum state preparation via stochastic prediction of neural network}}
\date{\today }

\begin{abstract}
In pursuit of enhancing the predication capabilities of the neural network, it has been a longstanding objective to create dataset encompassing a diverse array of samples. The purpose is to broaden the horizons of neural network and continually strive for improved prediction accuracy during training process, which serves as the ultimate evaluation metric. In this paper, we explore an intriguing avenue for enhancing algorithm effectiveness through exploiting the knowledge blindness of neural network. Our approach centers around a machine learning algorithm utilized for preparing arbitrary quantum states in a semiconductor double quantum dot system, a system characterized by highly constrained control degrees of freedom. By leveraging stochastic prediction generated by the neural network, we are able to guide the optimization process to escape local optima. Notably, unlike previous methodologies that employ reinforcement learning to identify pulse patterns, we adopt a training approach akin to supervised learning, ultimately using it to dynamically design the pulse sequence. This approach not only streamlines the learning process but also constrains the size of neural network, thereby improving the efficiency of algorithm.

\end{abstract}

\author{Chao-Chao Li, Run-Hong He, Zhao-Ming Wang\footnote{Corresponding author: wangzhaoming@ouc.edu.cn}}
\affiliation{College of Physics and Optoelectronic Engineering,Ocean University of China, Qingdao 266100, China}
\maketitle

\section{Introduction}

Robust quantum control is crucial for effective quantum computation and quantum information processing. The physical platform includes nuclear magnetic resonance experiments \cite{vandersypen2005nmr}, captured ions \cite{richerme2014non, yung2014transistor}, superconducting qubits \cite{devoret2013superconducting, wendin2017quantum}, nitrogen-vacancy centers \cite{childress2013diamond}, and semiconductor quantum dots \cite{zajac2018resonantly, huang2019fidelity, watson2018programmable, jang2020three, hanson2007spins, eriksson2004spin, zwanenburg2013silicon}. Among these, spin qubits in semiconductor quantum dots show promise due to their scalability and long coherence times \cite{kim2014quantum, kawakami2016gate, muhonen2014storing, maune2012coherent, bluhm2011dephasing, barthel2010interlaced, pla2013high}. The singlet-triplet ($S$-$T_{0}$) qubit, which is encoded in the singlet-triplet spin subspace of two electrons trapped in a double quantum dot (DQD), is widely used. The advantage over other qubit candidates includes fast qubit operation and independence from uniform fluctuations in the magnetic field, allowing for complete control by electrical pulses \cite{wang2014robust, taylor2005fault, wu2014two, nichol2017high}.
\par
Universal quantum computing relies on two-qubit gates capable of performing entanglement around different axes of the Bloch sphere and precise single-qubit rotations \cite{nielsen2002quantum}. Efficient and precise quantum gates control constructed by deep reinforcement learning has been investigated \cite{an2019deep}, which involves executing gates within the constraints of platform and mitigating errors during execution \cite{throckmorton2017fast}. In the case of singlet-triplet spin qubits in semiconductor DQD, fast electrical control of the exchange coupling is necessary for precise control of the rotation rate around the z-axis of the Bloch sphere \cite{wang2012composite}. The quantum state preparation (QSP) algorithm is commonly employed as a subroutine for various tasks. In particular, Refs. \cite{pinto2023simulation, zanetti2023simulating} utilize QSP to implement general quantum measurements on quantum systems and to simulate noisy quantum channels, respectively.

\par
Typically, performing an arbitrary quantum spin gate requires numerically solving a set of coupled nonlinear equations to determine a composite pulse sequence \cite{throckmorton2017fast, wang2014robust, wang2012composite}, which is resource and time-consuming. Machine learning, a field born out of artificial intelligence, enables the analysis of vast amounts of data beyond human ability or previously imagined methods of enumeration \cite{yang2018neural, heaton2018ian}, and has shown wide applicability on quantum control \cite{zhang2018automatic, yang2020optimizing, lin2020quantum, bukov2018reinforcement, kong2020artificial, palmieri2020experimental, wang2020deep, niu2019universal, gratsea2020universal, ma2022curriculum}. It is now an active research area and has demonstrated great success in solving lots of physical problems \cite{jordan2015machine, silver2016mastering}. Ref. \cite{yang2018neural} uses supervised learning algorithm to design pulse sequences that closely match solutions of nonlinear equations. However, in practice complex pulse shapes and lengthy execution times limit the application \cite{zhang2019does}. Refs. \cite{zhang2019does, he2021deep} utilize deep reinforcement learning \cite{an2019deep, niu2019universal, lin2020quantum, wang2020deep} to design discrete dynamic pulses for driving an initial state to a fixed state or resetting an arbitrary quantum state to a specific target state. In addition, deep reinforcement learning has successfully generated arbitrary states from specific states in nitrogen-vacancy center systems \cite{haug2020classifying}. By combining Refs. \cite{he2021deep, haug2020classifying}, driving between arbitrary quantum states can be realized.
\par

For the pulse design, several optimization methods are available and have been widely used, such as greedy algorithm (GA) \cite{cormen2022introduction, balaman2019chapter},  gradient ascent pulse engineering (GRAPE) \cite{khaneja2005optimal, rowland2012implementing}, and chopped random-basis optimization (CRAB) \cite{doria2011optimal, caneva2011chopped}. These traditional methods have proven to be effective for the optimal control of lots of quantum systems. However, a major challenge with these methods is that they often converge to local optima instead of global maxima. Consequently, the search may become stuck on a local maximum, leading to an insufficient fidelity. Ref.  \cite{he2021universal} overcomes this limitation by using a revised greedy (RG) algorithm to implement a universal quantum state preparation with a high fidelity, but it is less efficient because of trial and error at every step.
\par

In this paper, we propose a stochastic prediction (SP) of neural network strategy, which can obtain reliable pulse sequences for high fidelity universal quantum state preparation. We use a large number of initial and target states to train the neural network and subsequently use the well-trained network to generate the pulse sequence, providing the control trajectory for state preparation. Our dataset solely consists of non-local optima, following the definition provided in the Ref. \cite{he2021universal}. A local optimum is defined as a scenario where the fidelity fails to improve compared to the previous step, indicating that the network has reached a local maximum. During the state preparation process, we employ the knowledge blindness of the neural network to escape these local optima. When the network encounters such a situation for the first time, it randomly predicts an action, allowing us to break free from the local optimum. Concurrently, we employ a supervised learning algorithm that dynamically determines the control pulse at each step. This approach simplifies the learning process and enhances the efficiency of algorithm when compared to other methods. Our evaluation results show that our pulse design scheme is more efficient than traditional optimization methods in a discrete control space and higher fidelity can be obtained. Compared to conventional pulse optimization methods, our scheme jumps out of the local optimum via the network of randomly predicted pulses, while improving the preparation efficiency.

\section{model}
Semiconductor quantum dots are a promising candidate for quantum computing due to the advantage that it can be fully electrically driven \cite{zhang2018qubits}. Here, we describe the single-qubit and two-qubit models in $S$-$T_{0}$. The effective Hamiltonian of a single $S$-$T_{0}$ qubit controlled by an external electrical pulse is \cite{petta2005coherent, maune2012coherent, levy2002universal, malinowski2017notch, foletti2009universal}
\begin{equation}
H=J\sigma_z+h\sigma_x,
\label{eq:1}
\end{equation}
under the computational basis states: spin singlet state $|0\rangle=|S\rangle = (|\uparrow\downarrow\rangle-|\downarrow\uparrow\rangle)/\sqrt{2}$, and spin triplet state $|1\rangle=|T_{0}\rangle = (|\uparrow\downarrow\rangle+|\downarrow\uparrow\rangle)/\sqrt{2}$. 
$h$ is the Zeeman energy gap caused by magnetic field and it represents rotation around the x-axis of Bloch sphere. The exchange interaction $J$ causes rotation around the z-axis. $h$ is not easy to be changed experimentally, we assume it to be a constant $h=1$ \cite{wu2014two}. The reduced Planck constant $\hbar=1$ is assumed for simplicity throughout. Thus, the only controllable parameter is the exchange interaction $J$ between the two electrons, which determines the rate of rotation around the z-axis and can be adjusted by applying an external voltage. Due to the nature of the exchange coupling, $J$ is finite and non-negative \cite{zhang2019semiconductor}, and these constraints allow the construction of composite pulses for the implementation of universal quantum gates.

Quantum information processing typically requires a two-qubit entanglement gate. In semiconductor DQD, the Hamiltonian describing two entangled qubits based on Coulomb interactions can be expressed as follows \cite{haug2020classifying, shulman2012demonstration, taylor2005fault, nichol2017high, wang2015improving, van2011charge}
\begin{small} 
\begin{equation}
\begin{split}
H_{2-qubit}=
&\frac{\hbar}{2}(J_{1}(\sigma_{z}\otimes I)+J_{2}(I\otimes \sigma_{z})+h_{1}(\sigma_{x}\otimes I) \\
&+h_{2}(I\otimes \sigma_{x})+\frac{J_{12}}{2}((\sigma_{z}-I)\otimes (\sigma_{z}-I))),
\label{eq:3}
\end{split}
\end{equation}
\end{small} 
under the basis states of $\{|SS\rangle,|ST_{0}\rangle,|T_{0}S\rangle,|T_{0}T_{0}\rangle\}$. $J_{i}$ and $h_{i}$ represent the exchange interaction and magnetic field gradient across the double quantum dot, respectively, with the subscripts $i=1, 2$ denoting the corresponding qubits. Experimentally, the coupling strength $J_{12}$ between the qubits is proportional to $J_{1}J_{2}$, where both $J_{i}$ values need to be positive. For simplicity, we set $J_{12} = J_{1}J_{2}/2$ and $h_{1}=h_{2}=1$ as in Ref.~\cite{he2021universal}. To manipulate this two-qubit system, it is only necessary to control the electrical pulses that adjust $J_{1}$ and $J_{2}$.

\section{methods}

Now our task is to design discrete control pulses that can drive one arbitrary state to another arbitrary state. The pulse sequences are generated by training neural network, and 
the control trajectory is set as a segmented constant function. To optimize various parameters of the neural network, such as weights and biases, supervised learning requires a large input data set. During the process of constructing composite pulses, the trained neural network can predict the appropriate pulse based on input that is not part of the training set. To reduce computational cost, the control pulses are discretized into segmented constant function \cite{rowland2012implementing}, with the maximum evolution time $T$ uniformly divided into $N$ segments and the pulse duration $dt$ set at $T/N$. The fidelity $F$, which quantifies the distance between the evolution state and the target state, is used to assess the quality of the state preparation. $F=|\langle S_{n}| S_{tar}\rangle|^2$, where $S_{n}$ denotes the evolution state at a time step of $n$, and $S_{tar}$ represents the target state.

Our approach includes several steps: First, we construct a dataset with a large number of initial and target states and their corresponding actions, which is then used to put into a neural network for training. During training, the weights and biases of the neurons are continuously adjusted to improve the predictions of network. The trained network is saved as a model for use in future. Secondly, we begin by setting the initialization time step to $step=0$ and feeding a pair of initial state $S_{init}$ and target state $S_{tar}$ into the network model to calculate the fidelity $F$ of the initial state. And we define it as the maximum fidelity $F_{max}$. After feature extraction in the fully connected layer, we obtain the output of the actions, which is a set of discrete actions output as a probability distribution under the activation function, and the sum of these probabilities is 1. We choose the best action $a_{k}=argmax(action)$, which represents the pulse strength $J(t)$. Using the current quantum state $S_{init}$ and the action obtained from the network prediction, we calculate the evolution state $S_{n}=exp(-iH(a_{k})dt)S_{init}$ and its corresponding fidelity $F_{n}$ for the next moment. Then we compare it with the previous maximun fidelity and select the larger of the two values as $F_{max}$. The evolution state $S_{n}$ is then fed into the network model as the new initial state with the target state at the time step $step=step+1$. We repeat this process until either the time step reaches the maximum step $N$ or the fidelity exceeds a satisfactory threshold. The control trajectory for the quantum state preparation consists of the sequence of actions predicted by the neural network, with this sequence representing the solution for obtaining the maximum value of fidelity. At last, the trained neural network can formulate appropriate control trajectories for quantum states in the test set or other states in the  Hilbert space.

\begin{figure}[htbp]
	\centerline{\includegraphics[width=1\columnwidth]{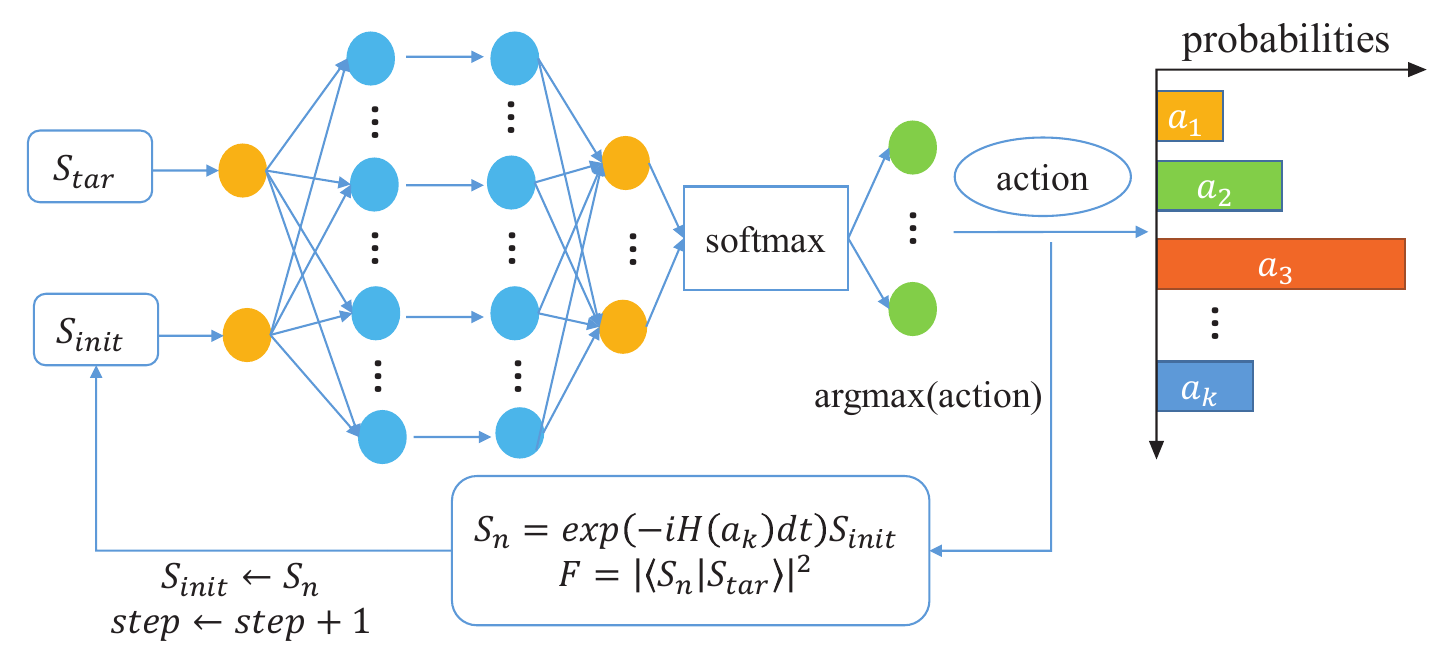}}
	\caption{Diagram of the stochastic prediction (SP) algorithm for designing control trajectory. The specifics of the algorithm are elaborated in section 3 and the pseudocode is presented in Algorithm 1.}
	\label{fig:1}
\end{figure}

\begin{algorithm}
	\caption{The pseudocode of the SP algorithm for designing control trajectory}
	\label{alg: AOA}
	\renewcommand{\algorithmicrequire}{\textbf{Input:}}
	\renewcommand{\algorithmicensure}{\textbf{Output:}}
	\begin{algorithmic}[1]
		\REQUIRE initial state $S_{init}$ and target state $S_{tar}$
		\ENSURE the maximum fidelity $F_{max}$ and pulse sequence 
		        from $step=0$ to $step_{end}$
		\STATE Calculate the initial fidelity $F$ and let $F_{max}=F$
		\STATE Initialize the time step $step=0$
		\WHILE{True}
		\STATE Feed initial and target states into the network model to predict the action probability distribution
		\STATE Choose the action $a_{k}=argmax(action)$
		\STATE Next state $S_{step}$ are the state obtained by performing $a_{k}$ and calculate the corresponding fidelity   
		       $F_{step}$
		\STATE Compare the two fidelities $F_{step}$ and $F_{max}$
		\IF {$F_{step}>F_{max}$}
		\STATE Let $F_{max}\leftarrow F_{step}$
		\ENDIF
		\STATE Let $S_{init}\leftarrow S_{step}$ and $step=step+1$
		\STATE Break if $F_{max}>0.999$ or $step>step_{max}$
		\ENDWHILE
	\end{algorithmic}
\end{algorithm}

During the dataset construction process, we exclude data points that correspond to local optima. This is determined by comparing the fidelity achieved in the subsequent step with that of the previous step. Specifically, if the fidelity does not improve, it remains unchanged or decreases, the data is deemed to be associated with a local optimum and is discarded. Consequently, the dataset solely comprises non-local optima, ensuring that the fidelity improves at each step. When the trained model is applied, the neural network exhibits the ability to escape local optima when encountering states trapped within them. Since the network has not been trained in similar scenarios, it randomly predicts the action to be performed, thereby generating a perturbation.
\par

Our work effectively tackles the dynamic decision-making problem by employing the widely utilized model of supervised learning in the field of machine learning. Within this framework, neural network are leveraged to determine action to be taken next, taking into account the current state.
In comparison to the RG algorithm, which explores suitable actions through trial and error, our approach offers a more direct and efficient means of obtaining the next action. By directly feeding the quantum state into the network model, our method proves to be more straightforward and highly efficient.
\par

The pulse design process is illustrated in Fig.~\ref{fig:1}. Algorithm.~\ref{alg: AOA} presents the pseudocode for the SP algorithm. The key feature of the SP algorithm is its randomness of prediction, which does not require the human intervention. Although the neural network may occasionally predict suboptimal actions that could result in a fidelity decrease from one step to the next, this global approach to state preparation can ultimately produce better results, avoiding the local optimality problem encountered by the traditional algorithms.

\section{Results and discussions}
In this section, we focus on the state preparation of single-qubit and two-qubit in semiconductor DQD and compare our approach with conventional optimization methods. The details of the default parameters of the algorithm are listed in Table.~\ref{table:1}.

\begin{table}[h!]
	\centering
	\caption{Default parameters of the neural network.}
	\label{table:1}
	\begin{tabular}{ccc} 
		\hline
		\textbf{Parameters} & \textbf{single-qubit} & \textbf{two-qubit}\\
		\hline
		Total evolution time & $4\pi$ & $10\pi$\\
		Action duration & $\pi/5$ & $\pi/2$\\
		Maximum time step & 20 & 20\\
		Number of allowed actions & 8 & 16\\
		Batch size & 64 & 128\\
		Neurons per hidden layer & 256/64/32/32/8 & 256/128/64/16\\
		Learning rate & 0.0005 & 0.001\\
		Number of epoch & 200 & 100\\
		Activation function & softmax & softmax\\
		\hline
	\end{tabular}
\end{table}

\subsection{Universal single-qubit state preparation}
An arbitrary single-qubit state can be represented by a point on the Bloch sphere $|\psi(\theta,\varphi)\rangle=\cos(\frac{\theta }{2})|0\rangle+e^{i\varphi}\sin(\frac{\theta }{2})|1\rangle$, where the polar angle $\theta\in[0,\pi]$ and the azimuthal angle $\varphi\in[0,2\pi)$. We take the dataset for a single-qubit state preparation as in Ref.~ \cite{he2021universal}, where 128 testing points distributed uniformly at the angles $\theta$ and $\phi$ are sampled on the Bloch sphere. Each of these points is prepared in turn as a target state, enabling us to assess the performance of our method. For one preparation task, there is one fidelity $\bar{F} $. The mean of these average fidelity $\langle \bar{F} \rangle$ is calculated over all target states. For example, the single-qubit state preparation corresponds to $128 * 127=16256$ tasks.  
 
\par
The state preparation can be achieved by performing successive rotations on a Bloch sphere, with the exchange coupling $J(t)$ as the only adjustable parameter \cite{throckmorton2017fast}.  In our approach, we use 8 discrete control pulses, $J\in\left\{0,1,2,3,4,5,6,7\right\}$. The total evolution time $T$ is set to $4\pi$, and the pulse duration $dt$ is set to $\pi/5$, resulting in a maximum allowed time step of $N=T/dt=20$ for the entire process. These parameters can be adjusted as required. 
\par

\begin{figure}[]
	\centerline{\includegraphics[width=0.8\columnwidth]{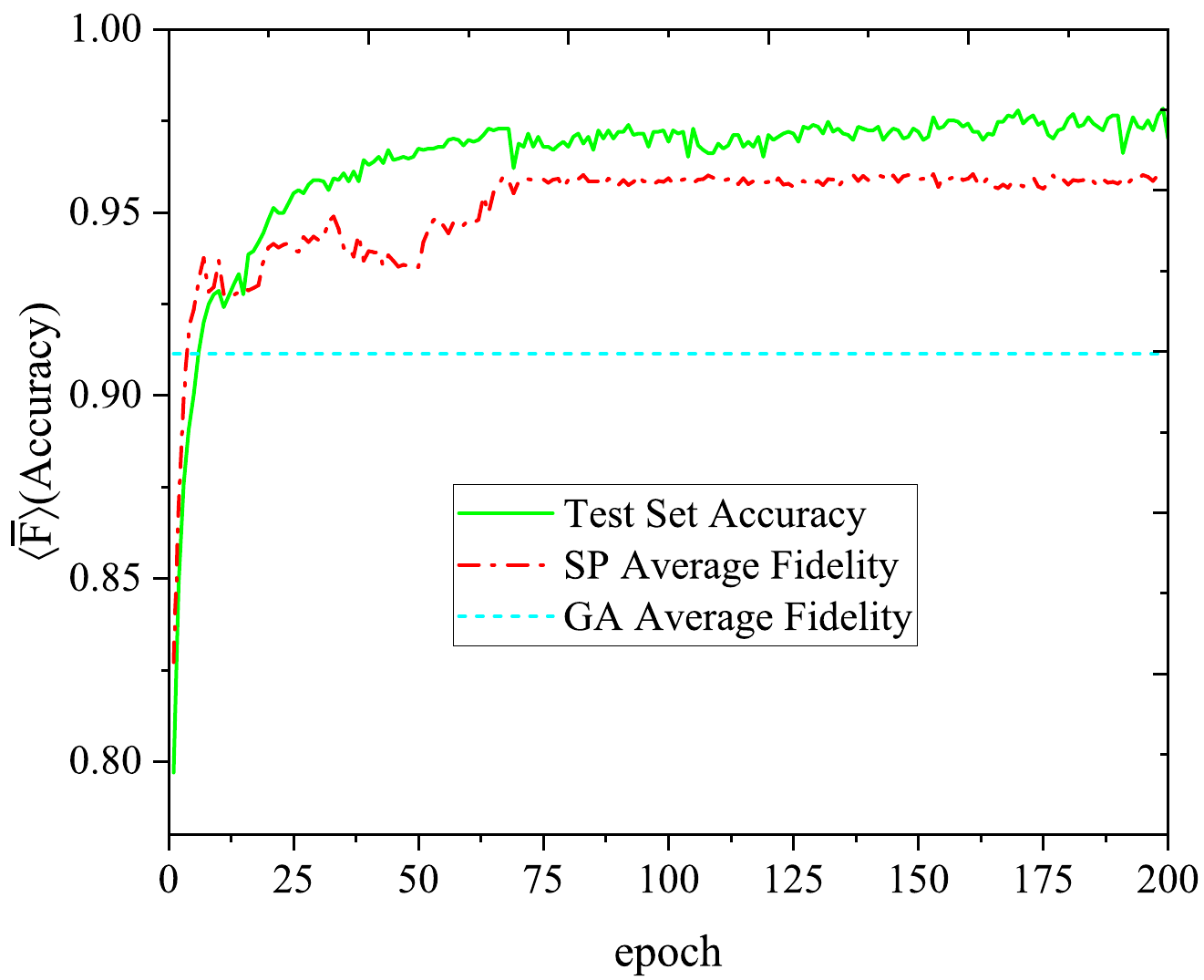}}
	\caption{The mean of all average fidelities $\langle \bar{F} \rangle$ of two algorithms and test set accuracy as the functions of the number of epoch in the training process for single-qubit preparation.}
	\label{fig:2}
\end{figure}

Fig.~\ref{fig:2} plots the test set accuracy, the average fidelity of the SP algorithm, and the average fidelity of the greedy algorithm versus the number of epoch during the neural network training. Classical algorithms such as the greedy algorithm are not involved in the training process, so the average fidelity of GA algorithm is a constant. Fig.~\ref{fig:2} shows that after about 75 epochs, the test set accuracy of the network and the average fidelity of the SP algorithm do not improve significantly as the number of epoch increases, indicating that the network has converged and the two trends are consistent. Furthermore, the average fidelity of the SP algorithm is significantly better than that of the greedy algorithm, demonstrating that our proposed scheme for preparing quantum states produces higher-quality results. Therefore, we conclude that our approach is viable, and the trained network can be applied to universal quantum state preparation tasks.
\par

\begin{figure}[]
	\centerline{\includegraphics[width=0.8\columnwidth]{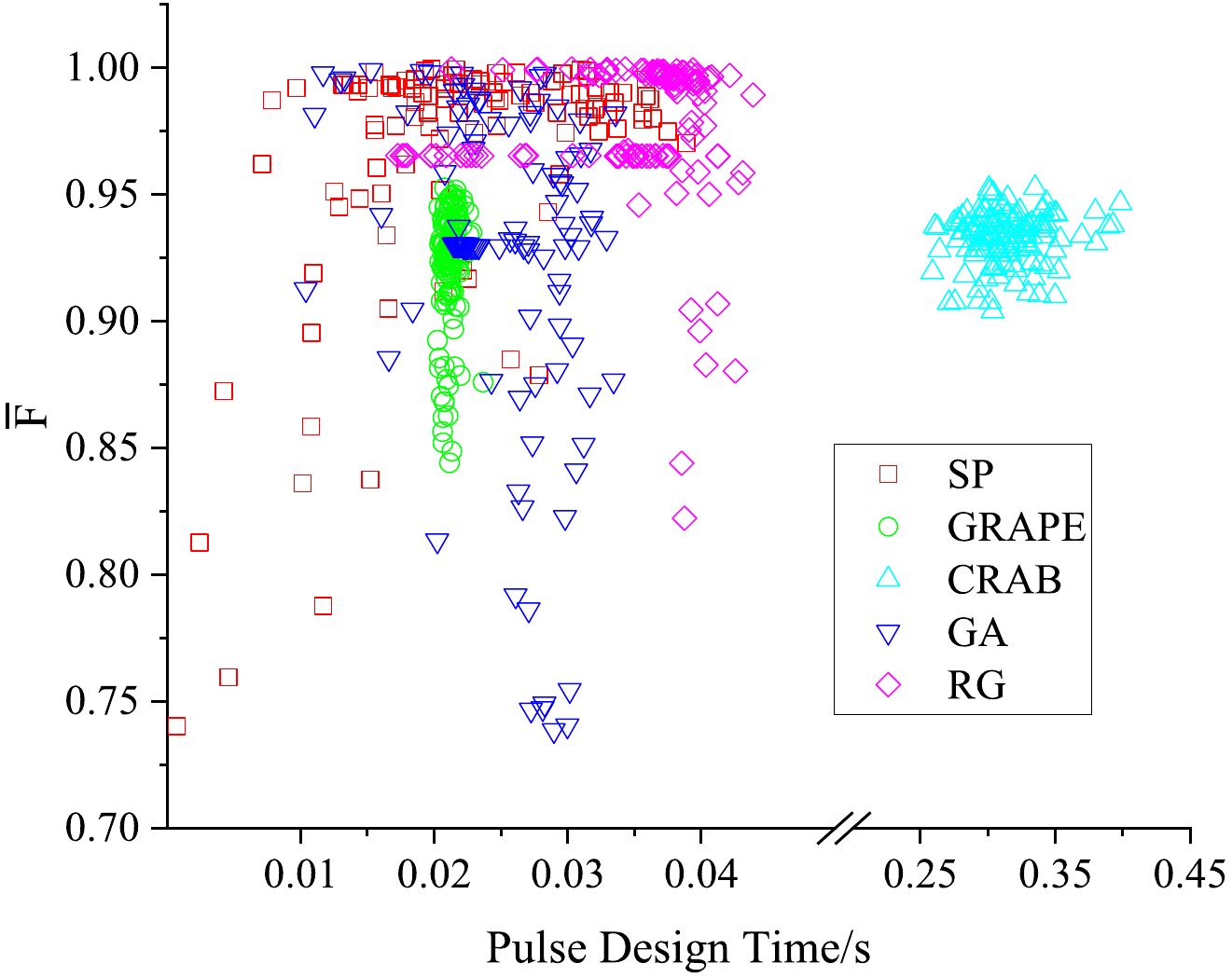}}
	\caption{The distribution of average fidelities $\bar{F}$ versus average design time $\bar{t}$ for the preparation of arbitrary single-qubit target states using various optimization algorithms, based on 128 sampled tasks. $\langle \bar{F} \rangle$ = 0.97, 0.9121, 0.9117, 0.9206, 0.97 and $\langle \bar{t} \rangle$ = 0.0211, 0.0212, 0.3142, 0.0246, 0.0347 with SP, GRAPE, CRAB, GA and RG, respectively. $\langle \bar{F} \rangle$ and $\langle \bar{t} \rangle$ represent the mean of all average fidelities and all average pulse design time over 128 preparation tasks.}
	\label{fig:3}
\end{figure}

High efficiency quantum state preparation means high fidelity with short design time. To evaluate the efficiency of our SP algorithm against other methods. We present 
the distribution of the average fidelity $\bar{F}$ versus the average designing time $\bar{t}$ of the SP, GRAPE, CRAB, GA and RG for preparation target states in Fig.~\ref{fig:3}. The control parameters are taken as the same as in Fig.~\ref{fig:2}. The average is based on the 128 state preparation tasks. To satisfy the discrete control requirement, we discretize the continuous control of GRAPE and CRAB to the nearest allowable action at the end of the execution \cite{zhang2019does}. As shown in Fig.~\ref{fig:3}, our SP algorithm outperforms all the other four conventional optimization algorithms in terms of efficiency in the discrete control space, with GRAPE and CRAB algorithms performing poorly in the same space. During optimization, the SP algorithm reduces the required time step adaptively to efficiently find the optimal solution. In contrast, GRAPE and CRAB use a fixed number of time steps and sometimes the optimal solution is missed. 

\subsection{Universal two-qubit state preparation}
For a two-qubit state preparation of a semiconductor DQD, the allowed control pulses for each qubit can be discretized as $\left\{(J_{1}, J_{2})|J_{1}, J_{2}\in\left\{1,2,3,4\right\}\right\}$, resulting in a total of 16 allowed actions. During this process, the total evolution time is set to $T=10\pi$ and the pulse duration to $dt=\pi/2$. The points in the data set for train and test are defined as $\left\{\left[a_{1}, a_{2}, a_{3}, a_{4}\right]^T\right\}$, where $a_{j}=e^{i\phi}c_{j}$ represents the probability amplitude of the corresponding $jth$ basis state, and $\phi\in\left\{0,\pi/2,\pi,3\pi/2\right\}$; and these $c_{j}s$ together represent the points on the hypersphere of the four-dimensional unit
\begin{equation}
\left\{
\begin{aligned}
	c_{1} &= \cos\theta_{1}, \\
	c_{2} &= \sin\theta_{1}\cos\theta_{2}, \\
	c_{3} &= \sin\theta_{1}\sin\theta_{2}\cos\theta_{3}, \\
	c_{4} &= \sin\theta_{1}\sin\theta_{2}\sin\theta_{3},
\end{aligned}
\right.
\label{eq:4}
\end{equation}
with $\theta_{i}\in\left\{\pi/8,\pi/4,3\pi/8\right\}$ \cite{he2021deep}. We select randomly 256 testing points to form the data set.

\begin{figure}[]
	\centerline{\includegraphics[width=0.8\columnwidth]{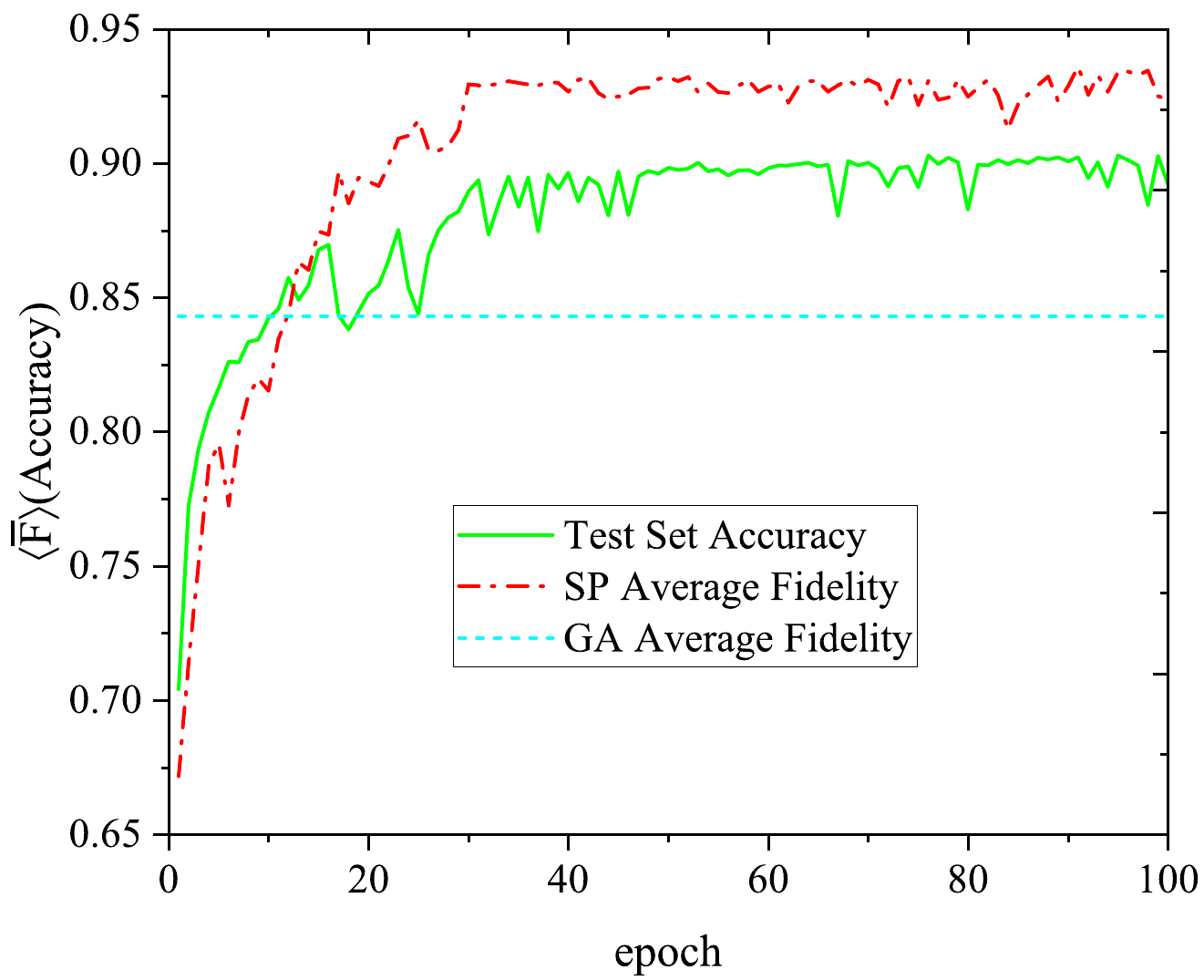}}
	\caption{The mean of the average fidelities $\langle \bar{F} \rangle$ of two algorithms and test set accuracy as the functions of the number of epoch in the training process for two-qubit preparation.}
	\label{fig:4}
\end{figure}

\begin{figure}[]
	\centerline{\includegraphics[width=0.8\columnwidth]{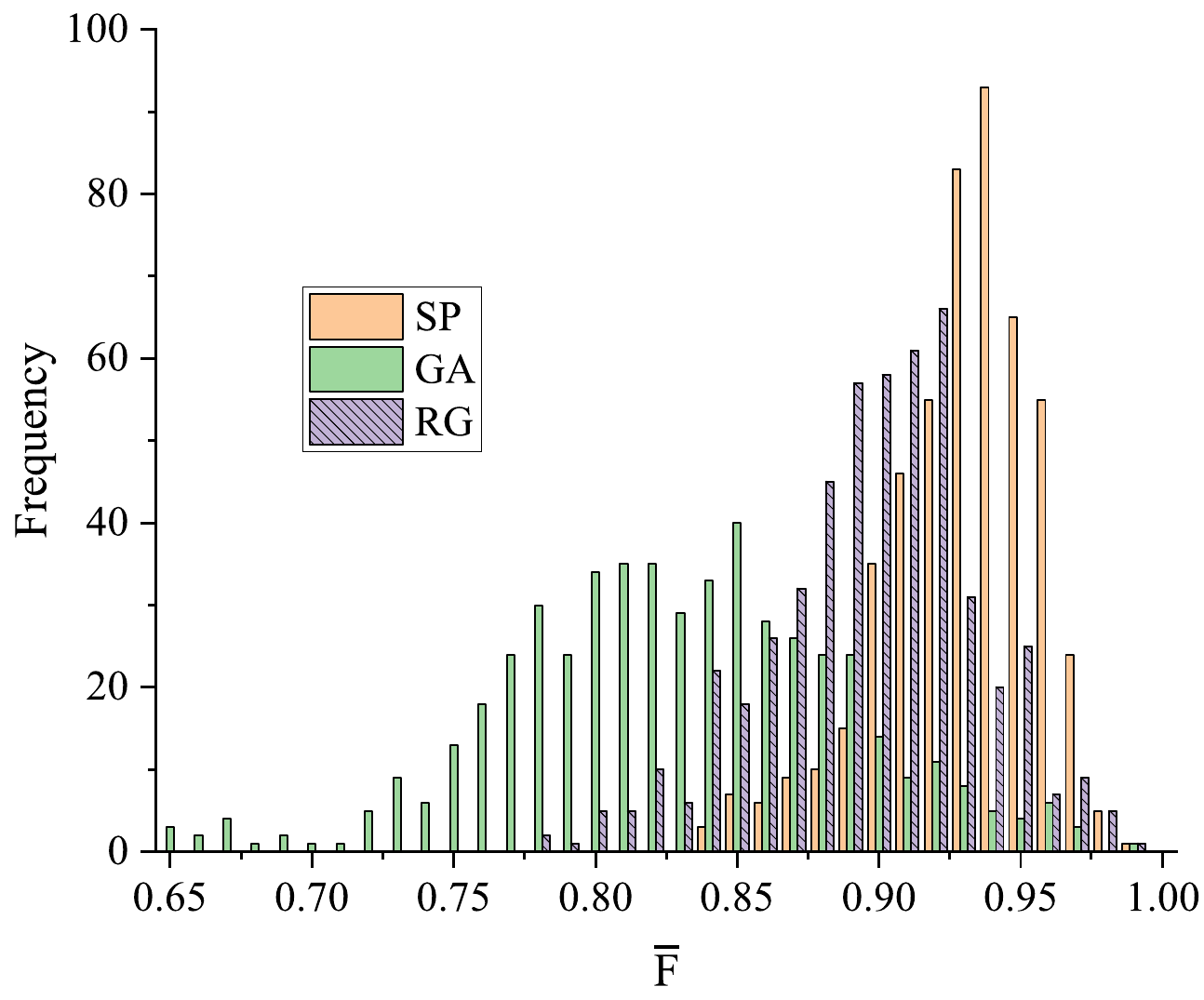}}
	\caption{The frequency distributions of average fidelities $\bar{F}$ for two-qubit preparation over 512 preparation tasks. The mean of all average fidelities $\langle \bar{F} \rangle$ = 0.9295, 0.8381, 0.8962 with SP, GA, RG.}
	\label{fig:5}
\end{figure}

As plotted in Fig.~\ref{fig:4}, the neural network converges after about 30 epochs. After 100 epochs of training, the average fidelity of the SP algorithm converges to 0.93. On the other hand, average fidelity of our proposed algorithm for the two-qubit state preparation still performs better than greedy algorithm. Fig.~\ref{fig:5} shows the frequency distribution of the average fidelity $\bar{F}$ for the 512 target states prepared by SP, GA and RG, respectively. The results again verify that our algorithm outperforms the other two algorithms. Although some bad spots exist, the overall performance is excellent.

\subsection{Universal state preparation in a noisy environment}
 
The complete quantumness is always expected when performing universal quantum state preparation. However, normally the quantum noise will destroy the quantumness and then decrease the fidelity. How the optimal pulse sequences perform when considering the noises? Next we introduce noises in the quantum line by adding the bit flip channel, phase flip channel, or amplitude damping channel, respectively. The bit flip and phase flip channels are modeled by applying an additional $X$ or $Z$ gate to the qubit with a probability of occurrence. These two noise channels are the so-called Pauli channels. We take the bit flip channel as an example, which can be expressed as
\begin{equation}
\epsilon(\rho)=(1-p)I\rho I+pX\rho X,
\label{eq:5}
\end{equation}
where $I$ is the unit matrix and $X$ is Pauli X gate. The corresponding Kraus operators for this channel are
\begin{equation}
E_{0}=\sqrt{1-p}\left[
\begin{array}{cc}
1 & 0 \\
0 & 1
\end{array}
\right],
E_{1}=\sqrt{p}\left[
\begin{array}{cc}
0 & 1 \\
1 & 0
\end{array}
\right],   
\label{eq:6}
\end{equation}
where $p$ is the probability of occurrence of bit flip. The amplitude damping channel accounts for the dissipation of energy from the quantum system and the mathematical form can be expressed as
\begin{equation}
\epsilon(\rho)=E_{0}\rho E_{0}^{\dagger}+E_{1}\rho E_{1}^{\dagger},
\label{eq:7}
\end{equation}
with Kraus operators
\begin{equation}
E_{0}=\left[
\begin{array}{cc}
1 & 0 \\
0 & \sqrt{1-p}
\end{array}
\right],
E_{1}=\left[
\begin{array}{cc}
0 & \sqrt{p} \\
0 & 0
\end{array}
\right],   
\label{eq:8}
\end{equation}
where $p$ is the dissipation factor. 
\begin{figure}
	\centering
	\subfigure[]{\includegraphics[width=0.8\columnwidth]{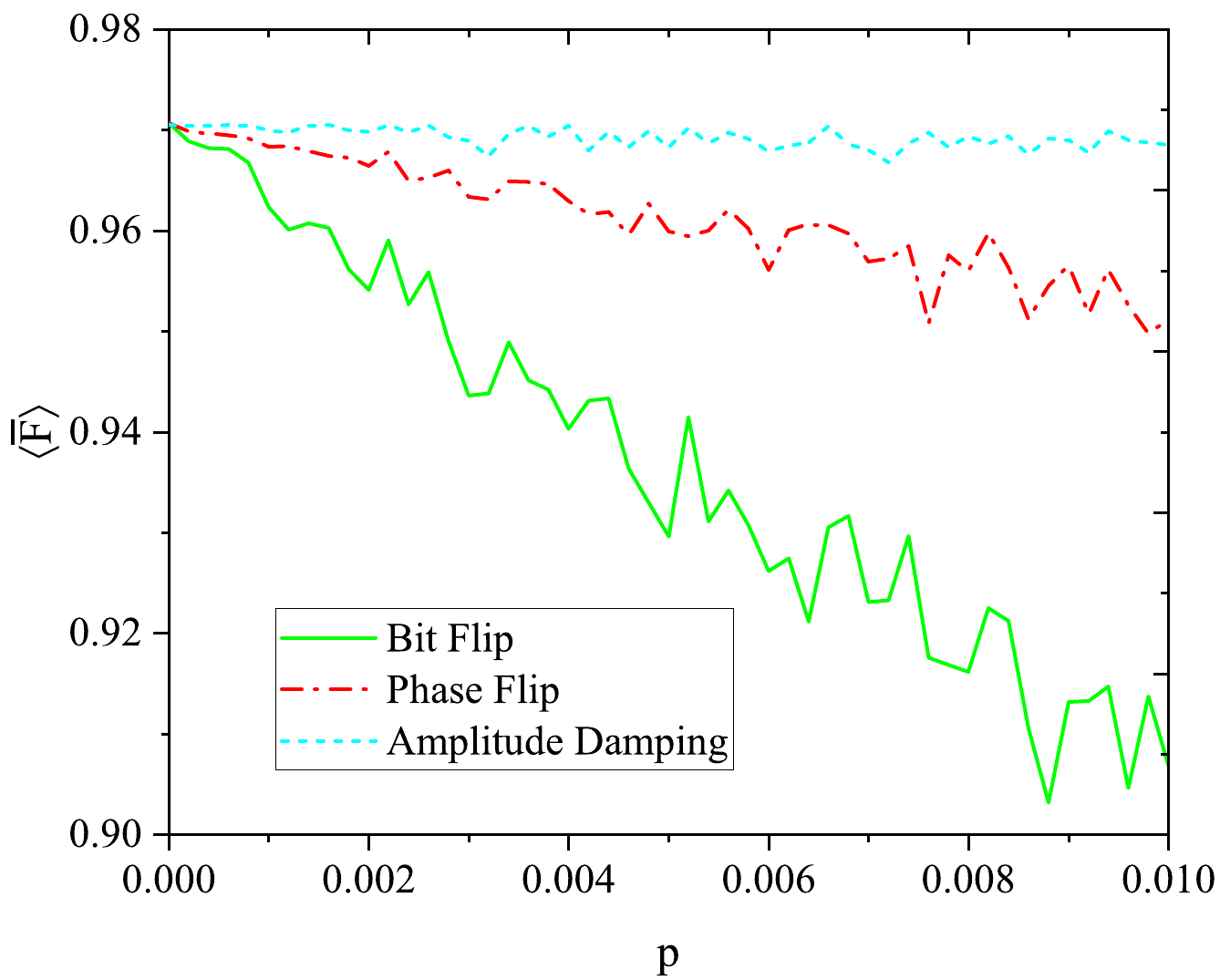}}
	\subfigure[]{\includegraphics[width=0.8\columnwidth]{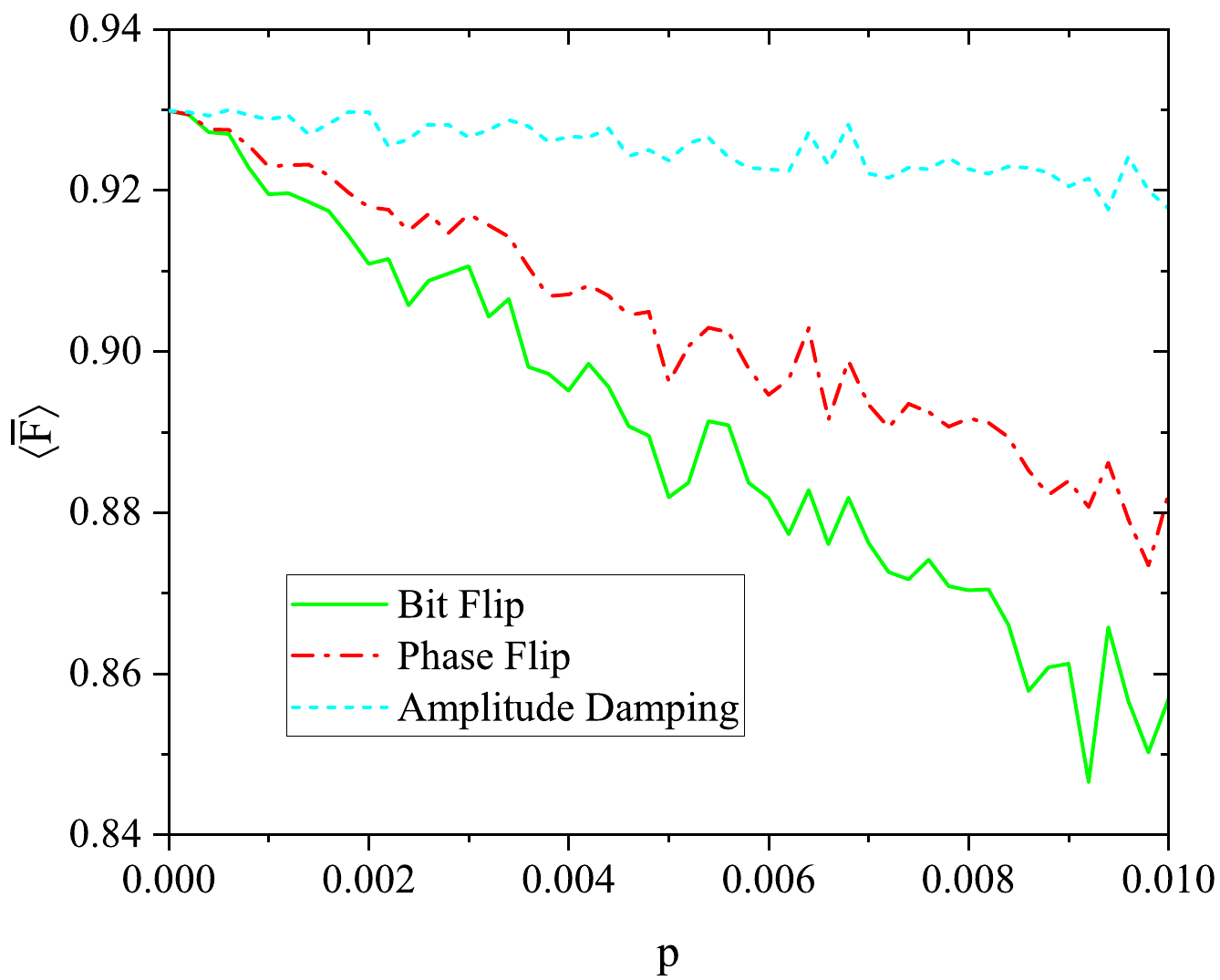}}
	\caption{The mean of average fidelities with SP algorithm versus occurrence probability or dissipation factor of different noise quantum channels. (a) single-qubit state preparation; (b) two-qubit state preparation.}
	\label{fig:6}
\end{figure}

We have found the ideal optimal pulse sequence that corresponds to the maximum fidelity through training the dataset in the absence of noises. For the noise model, we consider two cases. The first is that we use the ideal pulse sequence. We apply noise channel after each time step of the pulse to create a noisy quantum line, which is used to drive the initial state to the final state. For the second case, we directly search the optimal pulse in the noise model. 
Now for the first cases, in Fig.~\ref{fig:6} we plot the fidelity $\langle \bar{F} \rangle$ as a function of the occurrence probability (dissipation factor) $p$ for single-qubit and two-qubit state preparation. For two-qubit state preparation, we assume that both qubits noise channels are identical, and the probability or dissipation factor of the noise channels is the same ($p_{1}=p_{2}=p$). $\langle \bar{F} \rangle$ decreases with increasing $p$ as expected. For single and two qubit case and for the same $p$, $\langle \bar{F} \rangle$ decreases most significantly for the bit flip, phase flip is in the middle, and amplitude damping corresponds to the minimal impact.

\par

\begin{figure}
	\centering
	\subfigure[]{\includegraphics[width=0.75\columnwidth]{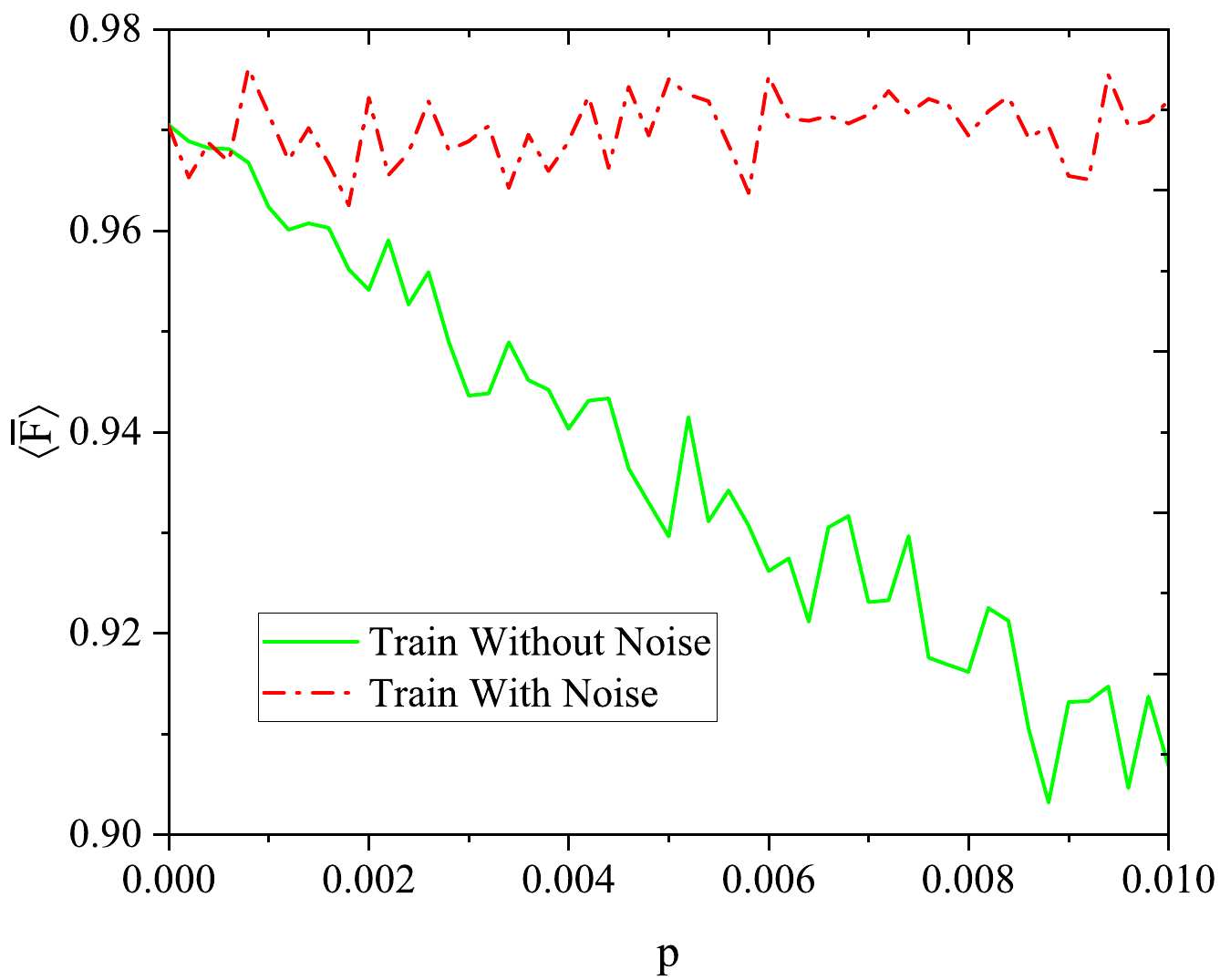}}
	\subfigure[]{\includegraphics[width=0.7\columnwidth]{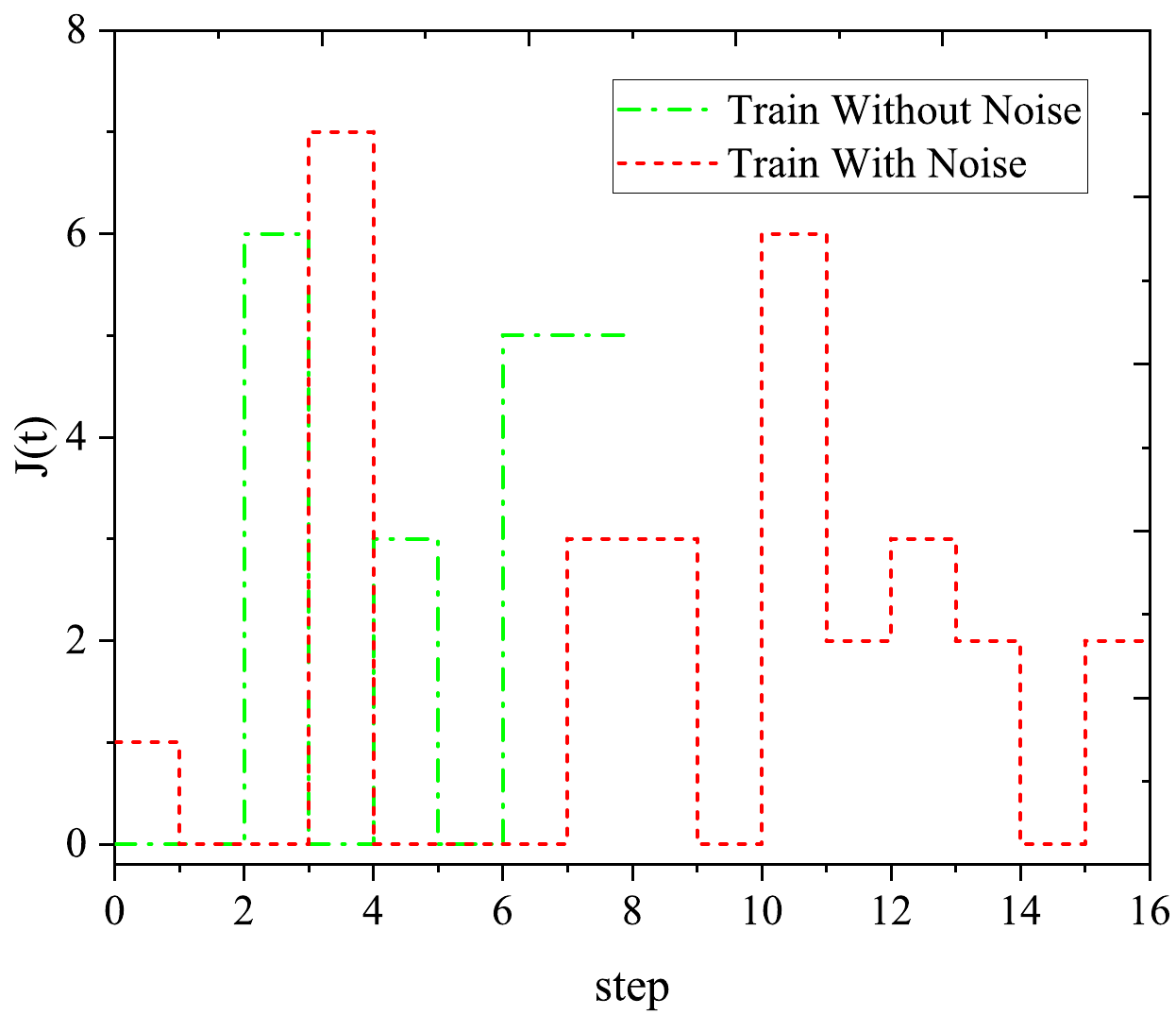}}
	\caption{(a) The mean of average fidelities using SP algorithm versus occurrence probability of bit flip channel with and without noise in train process; (b) The designed control trajectory for these two cases with occurrence probability $p=0.005$. The initial state and target state are set as $|0\rangle$ and $|1\rangle$.}
	\label{fig:7}
\end{figure}

\begin{figure}
	\centering
	\subfigure[]{\includegraphics[width=0.48\columnwidth]{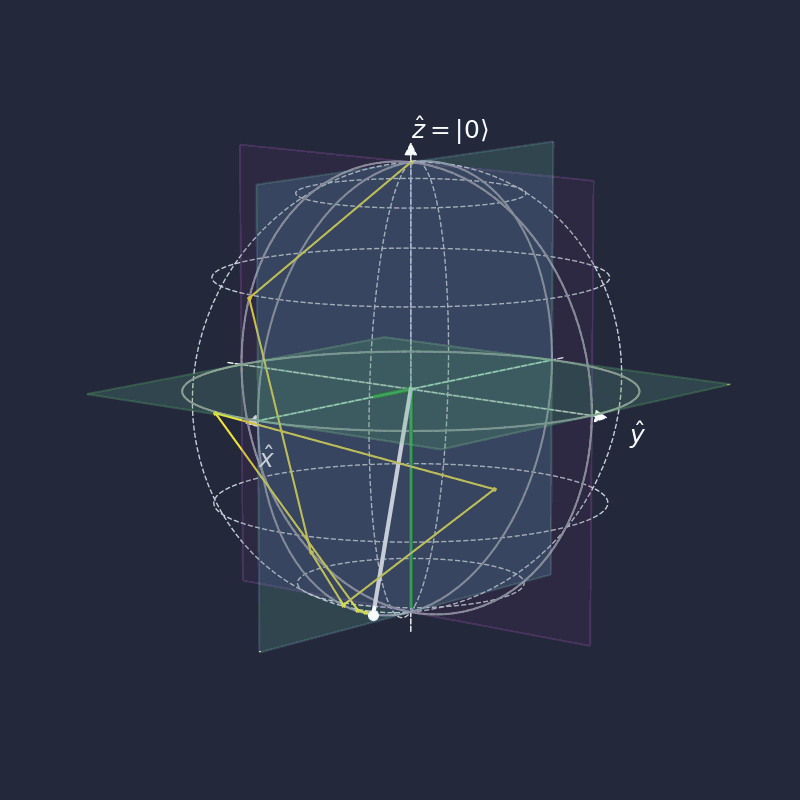}}
	\subfigure[]{\includegraphics[width=0.48\columnwidth]{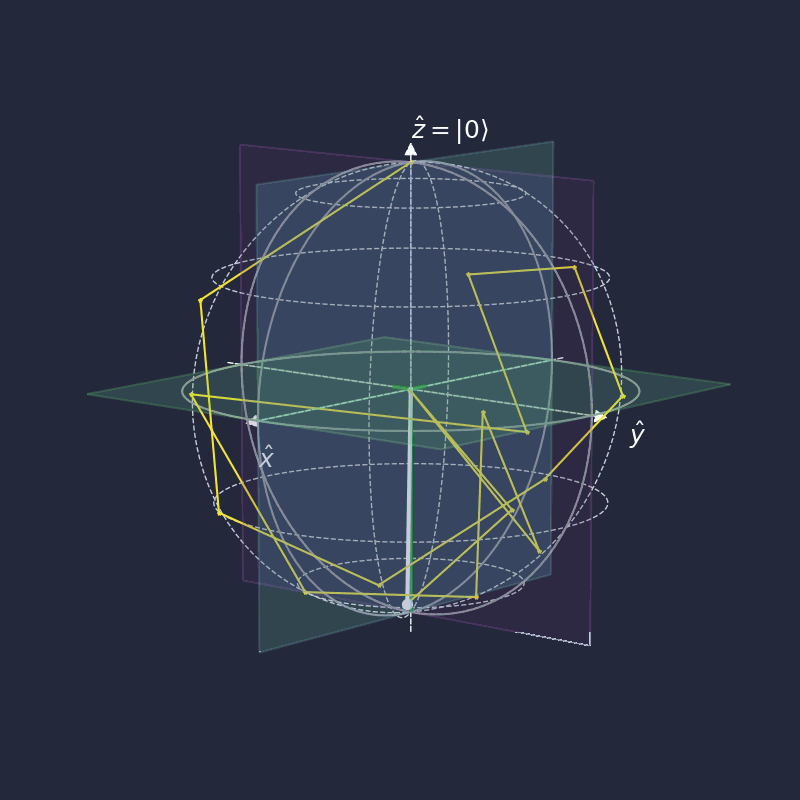}}
	\caption{The corresponding motion trail for the reset task on the Bloch sphere. (a) train without noise, the final fidelity F = 0.9752; (b) train with noise, the final fidelity F = 0.9858.}
	\label{fig:8}
\end{figure}

It has been clearly shown that the fidelity will decrease in the presence of noise even with the ideal pulse sequence. Can we use our method to directly design the pulse sequence with noises? In this cases, the optimal pulse sequence depends not only on the system but also the noises. In a recent paper \cite{yangyang}, optimized pulse sequences for the adiabatic speedup are found by using stochastic search procedures in a noisy environment. The detrimental effects of the environment on the system are reduced for the optimal pulses compared with the ideal closed-system pulses. Now we will train the network with noises. Initially, we add a noise channel on the Hamiltonian and construct a data set incorporating the noises. Then we obtain a model after training. For the demonstration, we take a single-qubit quantum state preparation with bit flip channel as an example. 
The initial state is taken as $|0\rangle$ and target state as $|1\rangle$. Fig.~\ref{fig:7}(a) plots the mean of the average fidelity versus the flip probability for these two cases: train with (without) noise. Obviously, when using the ideal pulse sequences, $\langle \bar{F} \rangle$ decreases quickly with increasing $p$ as expected. However, when training with noises, $\langle \bar{F} \rangle$ is almost stable for different $p$. The value of $\langle \bar{F} \rangle$ oscillates around 0.965. This result indicates that once the environmental parameter is given, the neural network is able to adjust its weight parameters appropriately during the training to combat the noises. Fig.~\ref{fig:7}(b) shows the obtained control trajectories using the SP algorithm training with and without noises. These two trajectories are different from both the steps and the strengths. It only needs 8 steps when training withour noise and 16 steps are required with noises. Furthermore, we plot the corresponding motion trail of Fig.~\ref{fig:7}(b) for the reset task from $|0\rangle$ to $|1\rangle$ on the Bloch sphere in Fig.~\ref{fig:8}.

\par
Given the limitations of available quantum computing, we simulate quantum computing on a classical computer and generate the corresponding data. Our algorithm was implemented using PYTHON 3.8.8, TensorFlow 2.12.0, QuTip 4.7.1, MindQuantum 0.8.0, running on a computer with a 6-core 2.60 GHz CPU and 16 GB of RAM.

\section{conclusions}
In this paper, we propose an efficient SP algorithm for designing control trajectories that can prepare an arbitrary state from an arbitrary state. The scheme involves training a large number of initial and target states along with their corresponding actions using neural network. Once the network is trained, it can be used to predict control pulses without any further training. We demonstrate that the efficacy of our approach on single-qubit and two-qubit of semiconductor quantum dot, highlighting its potential application in the future quantum computation. Our SP algorithm shows its advantage over conventional pulse optimization algorithms by achieving a higher fidelity. Furthermore, the control pulses are predicted directly through the network, resulting in shorter pulse design time than other numerical optimization algorithms. At last, we consider noises including Pauli channel and amplitude damping channel. We find that our SP algorithm is still effective for the design of pulse sequences when training with noises. Our investigation shows that the machine learning is a powerful tool for the design of control pulse sequences in quantum information processing.

\begin{acknowledgements}
This paper is based upon work supported by the Natural Science Foundation of Shandong Province (Grants No. ZR2021LLZ004) and Fundamental Research Funds for the Central Universities
(Grant No. 202364008).
\end{acknowledgements}

\bibliography{xianghan}

\begin{thebibliography}{64}
\expandafter\ifx\csname natexlab\endcsname\relax\def\natexlab#1{#1}\fi
\expandafter\ifx\csname bibnamefont\endcsname\relax
  \def\bibnamefont#1{#1}\fi
\expandafter\ifx\csname bibfnamefont\endcsname\relax
  \def\bibfnamefont#1{#1}\fi
\expandafter\ifx\csname citenamefont\endcsname\relax
  \def\citenamefont#1{#1}\fi
\expandafter\ifx\csname url\endcsname\relax
  \def\url#1{\texttt{#1}}\fi
\expandafter\ifx\csname urlprefix\endcsname\relax\def\urlprefix{URL }\fi
\providecommand{\bibinfo}[2]{#2}
\providecommand{\eprint}[2][]{\url{#2}}

\bibitem[{\citenamefont{Vandersypen and Chuang}(2005)}]{vandersypen2005nmr}
\bibinfo{author}{\bibfnamefont{L.~M.} \bibnamefont{Vandersypen}}
  \bibnamefont{and} \bibinfo{author}{\bibfnamefont{I.~L.}
  \bibnamefont{Chuang}}, \bibinfo{journal}{Reviews of modern physics}
  \textbf{\bibinfo{volume}{76}}, \bibinfo{pages}{1037} (\bibinfo{year}{2005}).

\bibitem[{\citenamefont{Richerme et~al.}(2014)\citenamefont{Richerme, Gong,
  Lee, Senko, Smith, Foss-Feig, Michalakis, Gorshkov, and
  Monroe}}]{richerme2014non}
\bibinfo{author}{\bibfnamefont{P.}~\bibnamefont{Richerme}},
  \bibinfo{author}{\bibfnamefont{Z.-X.} \bibnamefont{Gong}},
  \bibinfo{author}{\bibfnamefont{A.}~\bibnamefont{Lee}},
  \bibinfo{author}{\bibfnamefont{C.}~\bibnamefont{Senko}},
  \bibinfo{author}{\bibfnamefont{J.}~\bibnamefont{Smith}},
  \bibinfo{author}{\bibfnamefont{M.}~\bibnamefont{Foss-Feig}},
  \bibinfo{author}{\bibfnamefont{S.}~\bibnamefont{Michalakis}},
  \bibinfo{author}{\bibfnamefont{A.~V.} \bibnamefont{Gorshkov}},
  \bibnamefont{and} \bibinfo{author}{\bibfnamefont{C.}~\bibnamefont{Monroe}},
  \bibinfo{journal}{Nature} \textbf{\bibinfo{volume}{511}},
  \bibinfo{pages}{198} (\bibinfo{year}{2014}).

\bibitem[{\citenamefont{Yung et~al.}(2014)\citenamefont{Yung, Casanova,
  Mezzacapo, Mcclean, Lamata, Aspuru-Guzik, and Solano}}]{yung2014transistor}
\bibinfo{author}{\bibfnamefont{M.-H.} \bibnamefont{Yung}},
  \bibinfo{author}{\bibfnamefont{J.}~\bibnamefont{Casanova}},
  \bibinfo{author}{\bibfnamefont{A.}~\bibnamefont{Mezzacapo}},
  \bibinfo{author}{\bibfnamefont{J.}~\bibnamefont{Mcclean}},
  \bibinfo{author}{\bibfnamefont{L.}~\bibnamefont{Lamata}},
  \bibinfo{author}{\bibfnamefont{A.}~\bibnamefont{Aspuru-Guzik}},
  \bibnamefont{and} \bibinfo{author}{\bibfnamefont{E.}~\bibnamefont{Solano}},
  \bibinfo{journal}{Scientific reports} \textbf{\bibinfo{volume}{4}},
  \bibinfo{pages}{3589} (\bibinfo{year}{2014}).

\bibitem[{\citenamefont{Devoret and
  Schoelkopf}(2013)}]{devoret2013superconducting}
\bibinfo{author}{\bibfnamefont{M.~H.} \bibnamefont{Devoret}} \bibnamefont{and}
  \bibinfo{author}{\bibfnamefont{R.~J.} \bibnamefont{Schoelkopf}},
  \bibinfo{journal}{Science} \textbf{\bibinfo{volume}{339}},
  \bibinfo{pages}{1169} (\bibinfo{year}{2013}).

\bibitem[{\citenamefont{Wendin}(2017)}]{wendin2017quantum}
\bibinfo{author}{\bibfnamefont{G.}~\bibnamefont{Wendin}},
  \bibinfo{journal}{Reports on Progress in Physics}
  \textbf{\bibinfo{volume}{80}}, \bibinfo{pages}{106001}
  (\bibinfo{year}{2017}).

\bibitem[{\citenamefont{Childress and Hanson}(2013)}]{childress2013diamond}
\bibinfo{author}{\bibfnamefont{L.}~\bibnamefont{Childress}} \bibnamefont{and}
  \bibinfo{author}{\bibfnamefont{R.}~\bibnamefont{Hanson}},
  \bibinfo{journal}{MRS bulletin} \textbf{\bibinfo{volume}{38}},
  \bibinfo{pages}{134} (\bibinfo{year}{2013}).

\bibitem[{\citenamefont{Zajac et~al.}(2018)\citenamefont{Zajac, Sigillito,
  Russ, Borjans, Taylor, Burkard, and Petta}}]{zajac2018resonantly}
\bibinfo{author}{\bibfnamefont{D.~M.} \bibnamefont{Zajac}},
  \bibinfo{author}{\bibfnamefont{A.~J.} \bibnamefont{Sigillito}},
  \bibinfo{author}{\bibfnamefont{M.}~\bibnamefont{Russ}},
  \bibinfo{author}{\bibfnamefont{F.}~\bibnamefont{Borjans}},
  \bibinfo{author}{\bibfnamefont{J.~M.} \bibnamefont{Taylor}},
  \bibinfo{author}{\bibfnamefont{G.}~\bibnamefont{Burkard}}, \bibnamefont{and}
  \bibinfo{author}{\bibfnamefont{J.~R.} \bibnamefont{Petta}},
  \bibinfo{journal}{Science} \textbf{\bibinfo{volume}{359}},
  \bibinfo{pages}{439} (\bibinfo{year}{2018}).

\bibitem[{\citenamefont{Huang et~al.}(2019)\citenamefont{Huang, Yang, Chan,
  Tanttu, Hensen, Leon, Fogarty, Hwang, Hudson, Itoh
  et~al.}}]{huang2019fidelity}
\bibinfo{author}{\bibfnamefont{W.}~\bibnamefont{Huang}},
  \bibinfo{author}{\bibfnamefont{C.}~\bibnamefont{Yang}},
  \bibinfo{author}{\bibfnamefont{K.}~\bibnamefont{Chan}},
  \bibinfo{author}{\bibfnamefont{T.}~\bibnamefont{Tanttu}},
  \bibinfo{author}{\bibfnamefont{B.}~\bibnamefont{Hensen}},
  \bibinfo{author}{\bibfnamefont{R.}~\bibnamefont{Leon}},
  \bibinfo{author}{\bibfnamefont{M.}~\bibnamefont{Fogarty}},
  \bibinfo{author}{\bibfnamefont{J.}~\bibnamefont{Hwang}},
  \bibinfo{author}{\bibfnamefont{F.}~\bibnamefont{Hudson}},
  \bibinfo{author}{\bibfnamefont{K.~M.} \bibnamefont{Itoh}},
  \bibnamefont{et~al.}, \bibinfo{journal}{Nature}
  \textbf{\bibinfo{volume}{569}}, \bibinfo{pages}{532} (\bibinfo{year}{2019}).

\bibitem[{\citenamefont{Watson et~al.}(2018)\citenamefont{Watson, Philips,
  Kawakami, Ward, Scarlino, Veldhorst, Savage, Lagally, Friesen, Coppersmith
  et~al.}}]{watson2018programmable}
\bibinfo{author}{\bibfnamefont{T.}~\bibnamefont{Watson}},
  \bibinfo{author}{\bibfnamefont{S.}~\bibnamefont{Philips}},
  \bibinfo{author}{\bibfnamefont{E.}~\bibnamefont{Kawakami}},
  \bibinfo{author}{\bibfnamefont{D.}~\bibnamefont{Ward}},
  \bibinfo{author}{\bibfnamefont{P.}~\bibnamefont{Scarlino}},
  \bibinfo{author}{\bibfnamefont{M.}~\bibnamefont{Veldhorst}},
  \bibinfo{author}{\bibfnamefont{D.}~\bibnamefont{Savage}},
  \bibinfo{author}{\bibfnamefont{M.}~\bibnamefont{Lagally}},
  \bibinfo{author}{\bibfnamefont{M.}~\bibnamefont{Friesen}},
  \bibinfo{author}{\bibfnamefont{S.}~\bibnamefont{Coppersmith}},
  \bibnamefont{et~al.}, \bibinfo{journal}{nature}
  \textbf{\bibinfo{volume}{555}}, \bibinfo{pages}{633} (\bibinfo{year}{2018}).

\bibitem[{\citenamefont{Jang et~al.}(2020)\citenamefont{Jang, Cho, Kim, Chung,
  Umansky, and Kim}}]{jang2020three}
\bibinfo{author}{\bibfnamefont{W.}~\bibnamefont{Jang}},
  \bibinfo{author}{\bibfnamefont{M.-K.} \bibnamefont{Cho}},
  \bibinfo{author}{\bibfnamefont{J.}~\bibnamefont{Kim}},
  \bibinfo{author}{\bibfnamefont{H.}~\bibnamefont{Chung}},
  \bibinfo{author}{\bibfnamefont{V.}~\bibnamefont{Umansky}}, \bibnamefont{and}
  \bibinfo{author}{\bibfnamefont{D.}~\bibnamefont{Kim}},
  \bibinfo{journal}{arXiv preprint arXiv:2009.13182}  (\bibinfo{year}{2020}).

\bibitem[{\citenamefont{Hanson et~al.}(2007)\citenamefont{Hanson, Kouwenhoven,
  Petta, Tarucha, and Vandersypen}}]{hanson2007spins}
\bibinfo{author}{\bibfnamefont{R.}~\bibnamefont{Hanson}},
  \bibinfo{author}{\bibfnamefont{L.~P.} \bibnamefont{Kouwenhoven}},
  \bibinfo{author}{\bibfnamefont{J.~R.} \bibnamefont{Petta}},
  \bibinfo{author}{\bibfnamefont{S.}~\bibnamefont{Tarucha}}, \bibnamefont{and}
  \bibinfo{author}{\bibfnamefont{L.~M.} \bibnamefont{Vandersypen}},
  \bibinfo{journal}{Reviews of modern physics} \textbf{\bibinfo{volume}{79}},
  \bibinfo{pages}{1217} (\bibinfo{year}{2007}).

\bibitem[{\citenamefont{Eriksson et~al.}(2004)\citenamefont{Eriksson, Friesen,
  Coppersmith, Joynt, Klein, Slinker, Tahan, Mooney, Chu, and
  Koester}}]{eriksson2004spin}
\bibinfo{author}{\bibfnamefont{M.~A.} \bibnamefont{Eriksson}},
  \bibinfo{author}{\bibfnamefont{M.}~\bibnamefont{Friesen}},
  \bibinfo{author}{\bibfnamefont{S.~N.} \bibnamefont{Coppersmith}},
  \bibinfo{author}{\bibfnamefont{R.}~\bibnamefont{Joynt}},
  \bibinfo{author}{\bibfnamefont{L.~J.} \bibnamefont{Klein}},
  \bibinfo{author}{\bibfnamefont{K.}~\bibnamefont{Slinker}},
  \bibinfo{author}{\bibfnamefont{C.}~\bibnamefont{Tahan}},
  \bibinfo{author}{\bibfnamefont{P.}~\bibnamefont{Mooney}},
  \bibinfo{author}{\bibfnamefont{J.}~\bibnamefont{Chu}}, \bibnamefont{and}
  \bibinfo{author}{\bibfnamefont{S.}~\bibnamefont{Koester}},
  \bibinfo{journal}{Quantum Information Processing}
  \textbf{\bibinfo{volume}{3}}, \bibinfo{pages}{133} (\bibinfo{year}{2004}).

\bibitem[{\citenamefont{Zwanenburg et~al.}(2013)\citenamefont{Zwanenburg,
  Dzurak, Morello, Simmons, Hollenberg, Klimeck, Rogge, Coppersmith, and
  Eriksson}}]{zwanenburg2013silicon}
\bibinfo{author}{\bibfnamefont{F.~A.} \bibnamefont{Zwanenburg}},
  \bibinfo{author}{\bibfnamefont{A.~S.} \bibnamefont{Dzurak}},
  \bibinfo{author}{\bibfnamefont{A.}~\bibnamefont{Morello}},
  \bibinfo{author}{\bibfnamefont{M.~Y.} \bibnamefont{Simmons}},
  \bibinfo{author}{\bibfnamefont{L.~C.} \bibnamefont{Hollenberg}},
  \bibinfo{author}{\bibfnamefont{G.}~\bibnamefont{Klimeck}},
  \bibinfo{author}{\bibfnamefont{S.}~\bibnamefont{Rogge}},
  \bibinfo{author}{\bibfnamefont{S.~N.} \bibnamefont{Coppersmith}},
  \bibnamefont{and} \bibinfo{author}{\bibfnamefont{M.~A.}
  \bibnamefont{Eriksson}}, \bibinfo{journal}{Reviews of modern physics}
  \textbf{\bibinfo{volume}{85}}, \bibinfo{pages}{961} (\bibinfo{year}{2013}).

\bibitem[{\citenamefont{Kim et~al.}(2014)\citenamefont{Kim, Shi, Simmons, Ward,
  Prance, Koh, Gamble, Savage, Lagally, Friesen et~al.}}]{kim2014quantum}
\bibinfo{author}{\bibfnamefont{D.}~\bibnamefont{Kim}},
  \bibinfo{author}{\bibfnamefont{Z.}~\bibnamefont{Shi}},
  \bibinfo{author}{\bibfnamefont{C.}~\bibnamefont{Simmons}},
  \bibinfo{author}{\bibfnamefont{D.}~\bibnamefont{Ward}},
  \bibinfo{author}{\bibfnamefont{J.}~\bibnamefont{Prance}},
  \bibinfo{author}{\bibfnamefont{T.~S.} \bibnamefont{Koh}},
  \bibinfo{author}{\bibfnamefont{J.~K.} \bibnamefont{Gamble}},
  \bibinfo{author}{\bibfnamefont{D.}~\bibnamefont{Savage}},
  \bibinfo{author}{\bibfnamefont{M.}~\bibnamefont{Lagally}},
  \bibinfo{author}{\bibfnamefont{M.}~\bibnamefont{Friesen}},
  \bibnamefont{et~al.}, \bibinfo{journal}{Nature}
  \textbf{\bibinfo{volume}{511}}, \bibinfo{pages}{70} (\bibinfo{year}{2014}).

\bibitem[{\citenamefont{Kawakami et~al.}(2016)\citenamefont{Kawakami, Jullien,
  Scarlino, Ward, Savage, Lagally, Dobrovitski, Friesen, Coppersmith, Eriksson
  et~al.}}]{kawakami2016gate}
\bibinfo{author}{\bibfnamefont{E.}~\bibnamefont{Kawakami}},
  \bibinfo{author}{\bibfnamefont{T.}~\bibnamefont{Jullien}},
  \bibinfo{author}{\bibfnamefont{P.}~\bibnamefont{Scarlino}},
  \bibinfo{author}{\bibfnamefont{D.~R.} \bibnamefont{Ward}},
  \bibinfo{author}{\bibfnamefont{D.~E.} \bibnamefont{Savage}},
  \bibinfo{author}{\bibfnamefont{M.~G.} \bibnamefont{Lagally}},
  \bibinfo{author}{\bibfnamefont{V.~V.} \bibnamefont{Dobrovitski}},
  \bibinfo{author}{\bibfnamefont{M.}~\bibnamefont{Friesen}},
  \bibinfo{author}{\bibfnamefont{S.~N.} \bibnamefont{Coppersmith}},
  \bibinfo{author}{\bibfnamefont{M.~A.} \bibnamefont{Eriksson}},
  \bibnamefont{et~al.}, \bibinfo{journal}{Proceedings of the National Academy
  of Sciences} \textbf{\bibinfo{volume}{113}}, \bibinfo{pages}{11738}
  (\bibinfo{year}{2016}).

\bibitem[{\citenamefont{Muhonen et~al.}(2014)\citenamefont{Muhonen, Dehollain,
  Laucht, Hudson, Kalra, Sekiguchi, Itoh, Jamieson, McCallum, Dzurak
  et~al.}}]{muhonen2014storing}
\bibinfo{author}{\bibfnamefont{J.~T.} \bibnamefont{Muhonen}},
  \bibinfo{author}{\bibfnamefont{J.~P.} \bibnamefont{Dehollain}},
  \bibinfo{author}{\bibfnamefont{A.}~\bibnamefont{Laucht}},
  \bibinfo{author}{\bibfnamefont{F.~E.} \bibnamefont{Hudson}},
  \bibinfo{author}{\bibfnamefont{R.}~\bibnamefont{Kalra}},
  \bibinfo{author}{\bibfnamefont{T.}~\bibnamefont{Sekiguchi}},
  \bibinfo{author}{\bibfnamefont{K.~M.} \bibnamefont{Itoh}},
  \bibinfo{author}{\bibfnamefont{D.~N.} \bibnamefont{Jamieson}},
  \bibinfo{author}{\bibfnamefont{J.~C.} \bibnamefont{McCallum}},
  \bibinfo{author}{\bibfnamefont{A.~S.} \bibnamefont{Dzurak}},
  \bibnamefont{et~al.}, \bibinfo{journal}{Nature nanotechnology}
  \textbf{\bibinfo{volume}{9}}, \bibinfo{pages}{986} (\bibinfo{year}{2014}).

\bibitem[{\citenamefont{Maune et~al.}(2012)\citenamefont{Maune, Borselli,
  Huang, Ladd, Deelman, Holabird, Kiselev, Alvarado-Rodriguez, Ross, Schmitz
  et~al.}}]{maune2012coherent}
\bibinfo{author}{\bibfnamefont{B.~M.} \bibnamefont{Maune}},
  \bibinfo{author}{\bibfnamefont{M.~G.} \bibnamefont{Borselli}},
  \bibinfo{author}{\bibfnamefont{B.}~\bibnamefont{Huang}},
  \bibinfo{author}{\bibfnamefont{T.~D.} \bibnamefont{Ladd}},
  \bibinfo{author}{\bibfnamefont{P.~W.} \bibnamefont{Deelman}},
  \bibinfo{author}{\bibfnamefont{K.~S.} \bibnamefont{Holabird}},
  \bibinfo{author}{\bibfnamefont{A.~A.} \bibnamefont{Kiselev}},
  \bibinfo{author}{\bibfnamefont{I.}~\bibnamefont{Alvarado-Rodriguez}},
  \bibinfo{author}{\bibfnamefont{R.~S.} \bibnamefont{Ross}},
  \bibinfo{author}{\bibfnamefont{A.~E.} \bibnamefont{Schmitz}},
  \bibnamefont{et~al.}, \bibinfo{journal}{Nature}
  \textbf{\bibinfo{volume}{481}}, \bibinfo{pages}{344} (\bibinfo{year}{2012}).

\bibitem[{\citenamefont{Bluhm et~al.}(2011)\citenamefont{Bluhm, Foletti, Neder,
  Rudner, Mahalu, Umansky, and Yacoby}}]{bluhm2011dephasing}
\bibinfo{author}{\bibfnamefont{H.}~\bibnamefont{Bluhm}},
  \bibinfo{author}{\bibfnamefont{S.}~\bibnamefont{Foletti}},
  \bibinfo{author}{\bibfnamefont{I.}~\bibnamefont{Neder}},
  \bibinfo{author}{\bibfnamefont{M.}~\bibnamefont{Rudner}},
  \bibinfo{author}{\bibfnamefont{D.}~\bibnamefont{Mahalu}},
  \bibinfo{author}{\bibfnamefont{V.}~\bibnamefont{Umansky}}, \bibnamefont{and}
  \bibinfo{author}{\bibfnamefont{A.}~\bibnamefont{Yacoby}},
  \bibinfo{journal}{Nature Physics} \textbf{\bibinfo{volume}{7}},
  \bibinfo{pages}{109} (\bibinfo{year}{2011}).

\bibitem[{\citenamefont{Barthel et~al.}(2010)\citenamefont{Barthel, Medford,
  Marcus, Hanson, and Gossard}}]{barthel2010interlaced}
\bibinfo{author}{\bibfnamefont{C.}~\bibnamefont{Barthel}},
  \bibinfo{author}{\bibfnamefont{J.}~\bibnamefont{Medford}},
  \bibinfo{author}{\bibfnamefont{C.}~\bibnamefont{Marcus}},
  \bibinfo{author}{\bibfnamefont{M.}~\bibnamefont{Hanson}}, \bibnamefont{and}
  \bibinfo{author}{\bibfnamefont{A.}~\bibnamefont{Gossard}},
  \bibinfo{journal}{Physical review letters} \textbf{\bibinfo{volume}{105}},
  \bibinfo{pages}{266808} (\bibinfo{year}{2010}).

\bibitem[{\citenamefont{Pla et~al.}(2013)\citenamefont{Pla, Tan, Dehollain,
  Lim, Morton, Zwanenburg, Jamieson, Dzurak, and Morello}}]{pla2013high}
\bibinfo{author}{\bibfnamefont{J.~J.} \bibnamefont{Pla}},
  \bibinfo{author}{\bibfnamefont{K.~Y.} \bibnamefont{Tan}},
  \bibinfo{author}{\bibfnamefont{J.~P.} \bibnamefont{Dehollain}},
  \bibinfo{author}{\bibfnamefont{W.~H.} \bibnamefont{Lim}},
  \bibinfo{author}{\bibfnamefont{J.~J.} \bibnamefont{Morton}},
  \bibinfo{author}{\bibfnamefont{F.~A.} \bibnamefont{Zwanenburg}},
  \bibinfo{author}{\bibfnamefont{D.~N.} \bibnamefont{Jamieson}},
  \bibinfo{author}{\bibfnamefont{A.~S.} \bibnamefont{Dzurak}},
  \bibnamefont{and} \bibinfo{author}{\bibfnamefont{A.}~\bibnamefont{Morello}},
  \bibinfo{journal}{Nature} \textbf{\bibinfo{volume}{496}},
  \bibinfo{pages}{334} (\bibinfo{year}{2013}).

\bibitem[{\citenamefont{Wang et~al.}(2014)\citenamefont{Wang, Bishop, Barnes,
  Kestner, and Sarma}}]{wang2014robust}
\bibinfo{author}{\bibfnamefont{X.}~\bibnamefont{Wang}},
  \bibinfo{author}{\bibfnamefont{L.~S.} \bibnamefont{Bishop}},
  \bibinfo{author}{\bibfnamefont{E.}~\bibnamefont{Barnes}},
  \bibinfo{author}{\bibfnamefont{J.}~\bibnamefont{Kestner}}, \bibnamefont{and}
  \bibinfo{author}{\bibfnamefont{S.~D.} \bibnamefont{Sarma}},
  \bibinfo{journal}{Physical Review A} \textbf{\bibinfo{volume}{89}},
  \bibinfo{pages}{022310} (\bibinfo{year}{2014}).

\bibitem[{\citenamefont{Taylor et~al.}(2005)\citenamefont{Taylor, Engel,
  D{\"u}r, Yacoby, Marcus, Zoller, and Lukin}}]{taylor2005fault}
\bibinfo{author}{\bibfnamefont{J.}~\bibnamefont{Taylor}},
  \bibinfo{author}{\bibfnamefont{H.-A.} \bibnamefont{Engel}},
  \bibinfo{author}{\bibfnamefont{W.}~\bibnamefont{D{\"u}r}},
  \bibinfo{author}{\bibfnamefont{A.}~\bibnamefont{Yacoby}},
  \bibinfo{author}{\bibfnamefont{C.}~\bibnamefont{Marcus}},
  \bibinfo{author}{\bibfnamefont{P.}~\bibnamefont{Zoller}}, \bibnamefont{and}
  \bibinfo{author}{\bibfnamefont{M.}~\bibnamefont{Lukin}},
  \bibinfo{journal}{Nature Physics} \textbf{\bibinfo{volume}{1}},
  \bibinfo{pages}{177} (\bibinfo{year}{2005}).

\bibitem[{\citenamefont{Wu et~al.}(2014)\citenamefont{Wu, Ward, Prance, Kim,
  Gamble, Mohr, Shi, Savage, Lagally, Friesen et~al.}}]{wu2014two}
\bibinfo{author}{\bibfnamefont{X.}~\bibnamefont{Wu}},
  \bibinfo{author}{\bibfnamefont{D.~R.} \bibnamefont{Ward}},
  \bibinfo{author}{\bibfnamefont{J.}~\bibnamefont{Prance}},
  \bibinfo{author}{\bibfnamefont{D.}~\bibnamefont{Kim}},
  \bibinfo{author}{\bibfnamefont{J.~K.} \bibnamefont{Gamble}},
  \bibinfo{author}{\bibfnamefont{R.}~\bibnamefont{Mohr}},
  \bibinfo{author}{\bibfnamefont{Z.}~\bibnamefont{Shi}},
  \bibinfo{author}{\bibfnamefont{D.}~\bibnamefont{Savage}},
  \bibinfo{author}{\bibfnamefont{M.}~\bibnamefont{Lagally}},
  \bibinfo{author}{\bibfnamefont{M.}~\bibnamefont{Friesen}},
  \bibnamefont{et~al.}, \bibinfo{journal}{Proceedings of the National Academy
  of Sciences} \textbf{\bibinfo{volume}{111}}, \bibinfo{pages}{11938}
  (\bibinfo{year}{2014}).

\bibitem[{\citenamefont{Nichol et~al.}(2017)\citenamefont{Nichol, Orona,
  Harvey, Fallahi, Gardner, Manfra, and Yacoby}}]{nichol2017high}
\bibinfo{author}{\bibfnamefont{J.~M.} \bibnamefont{Nichol}},
  \bibinfo{author}{\bibfnamefont{L.~A.} \bibnamefont{Orona}},
  \bibinfo{author}{\bibfnamefont{S.~P.} \bibnamefont{Harvey}},
  \bibinfo{author}{\bibfnamefont{S.}~\bibnamefont{Fallahi}},
  \bibinfo{author}{\bibfnamefont{G.~C.} \bibnamefont{Gardner}},
  \bibinfo{author}{\bibfnamefont{M.~J.} \bibnamefont{Manfra}},
  \bibnamefont{and} \bibinfo{author}{\bibfnamefont{A.}~\bibnamefont{Yacoby}},
  \bibinfo{journal}{npj Quantum Information} \textbf{\bibinfo{volume}{3}},
  \bibinfo{pages}{3} (\bibinfo{year}{2017}).

\bibitem[{\citenamefont{Nielsen and Chuang}(2002)}]{nielsen2002quantum}
\bibinfo{author}{\bibfnamefont{M.~A.} \bibnamefont{Nielsen}} \bibnamefont{and}
  \bibinfo{author}{\bibfnamefont{I.}~\bibnamefont{Chuang}},
  \emph{\bibinfo{title}{Quantum computation and quantum information}}
  (\bibinfo{year}{2002}).

\bibitem[{\citenamefont{An and Zhou}(2019)}]{an2019deep}
\bibinfo{author}{\bibfnamefont{Z.}~\bibnamefont{An}} \bibnamefont{and}
  \bibinfo{author}{\bibfnamefont{D.}~\bibnamefont{Zhou}},
  \bibinfo{journal}{Europhysics Letters} \textbf{\bibinfo{volume}{126}},
  \bibinfo{pages}{60002} (\bibinfo{year}{2019}).

\bibitem[{\citenamefont{Throckmorton et~al.}(2017)\citenamefont{Throckmorton,
  Zhang, Yang, Wang, Barnes, and Sarma}}]{throckmorton2017fast}
\bibinfo{author}{\bibfnamefont{R.~E.} \bibnamefont{Throckmorton}},
  \bibinfo{author}{\bibfnamefont{C.}~\bibnamefont{Zhang}},
  \bibinfo{author}{\bibfnamefont{X.-C.} \bibnamefont{Yang}},
  \bibinfo{author}{\bibfnamefont{X.}~\bibnamefont{Wang}},
  \bibinfo{author}{\bibfnamefont{E.}~\bibnamefont{Barnes}}, \bibnamefont{and}
  \bibinfo{author}{\bibfnamefont{S.~D.} \bibnamefont{Sarma}},
  \bibinfo{journal}{Physical Review B} \textbf{\bibinfo{volume}{96}},
  \bibinfo{pages}{195424} (\bibinfo{year}{2017}).

\bibitem[{\citenamefont{Wang et~al.}(2012)\citenamefont{Wang, Bishop, Kestner,
  Barnes, Sun, and Das~Sarma}}]{wang2012composite}
\bibinfo{author}{\bibfnamefont{X.}~\bibnamefont{Wang}},
  \bibinfo{author}{\bibfnamefont{L.~S.} \bibnamefont{Bishop}},
  \bibinfo{author}{\bibfnamefont{J.}~\bibnamefont{Kestner}},
  \bibinfo{author}{\bibfnamefont{E.}~\bibnamefont{Barnes}},
  \bibinfo{author}{\bibfnamefont{K.}~\bibnamefont{Sun}}, \bibnamefont{and}
  \bibinfo{author}{\bibfnamefont{S.}~\bibnamefont{Das~Sarma}},
  \bibinfo{journal}{Nature communications} \textbf{\bibinfo{volume}{3}},
  \bibinfo{pages}{997} (\bibinfo{year}{2012}).

\bibitem[{\citenamefont{Pinto et~al.}(2023)\citenamefont{Pinto, Zanetti, Basso,
  and Maziero}}]{pinto2023simulation}
\bibinfo{author}{\bibfnamefont{D.~F.} \bibnamefont{Pinto}},
  \bibinfo{author}{\bibfnamefont{M.~S.} \bibnamefont{Zanetti}},
  \bibinfo{author}{\bibfnamefont{M.~L.} \bibnamefont{Basso}}, \bibnamefont{and}
  \bibinfo{author}{\bibfnamefont{J.}~\bibnamefont{Maziero}},
  \bibinfo{journal}{Physical Review A} \textbf{\bibinfo{volume}{107}},
  \bibinfo{pages}{022411} (\bibinfo{year}{2023}).

\bibitem[{\citenamefont{Zanetti et~al.}(2023)\citenamefont{Zanetti, Pinto,
  Basso, and Maziero}}]{zanetti2023simulating}
\bibinfo{author}{\bibfnamefont{M.~S.} \bibnamefont{Zanetti}},
  \bibinfo{author}{\bibfnamefont{D.~F.} \bibnamefont{Pinto}},
  \bibinfo{author}{\bibfnamefont{M.~L.} \bibnamefont{Basso}}, \bibnamefont{and}
  \bibinfo{author}{\bibfnamefont{J.}~\bibnamefont{Maziero}},
  \bibinfo{journal}{Journal of Physics B: Atomic, Molecular and Optical
  Physics} \textbf{\bibinfo{volume}{56}}, \bibinfo{pages}{115501}
  (\bibinfo{year}{2023}).

\bibitem[{\citenamefont{Yang et~al.}(2018)\citenamefont{Yang, Yung, and
  Wang}}]{yang2018neural}
\bibinfo{author}{\bibfnamefont{X.-C.} \bibnamefont{Yang}},
  \bibinfo{author}{\bibfnamefont{M.-H.} \bibnamefont{Yung}}, \bibnamefont{and}
  \bibinfo{author}{\bibfnamefont{X.}~\bibnamefont{Wang}},
  \bibinfo{journal}{Physical Review A} \textbf{\bibinfo{volume}{97}},
  \bibinfo{pages}{042324} (\bibinfo{year}{2018}).

\bibitem[{\citenamefont{Heaton}(2018)}]{heaton2018ian}
\bibinfo{author}{\bibfnamefont{J.}~\bibnamefont{Heaton}},
  \bibinfo{journal}{Genetic Programming and Evolvable Machines}
  \textbf{\bibinfo{volume}{19}}, \bibinfo{pages}{305} (\bibinfo{year}{2018}).

\bibitem[{\citenamefont{Zhang et~al.}(2018{\natexlab{a}})\citenamefont{Zhang,
  Cui, Wang, and Yung}}]{zhang2018automatic}
\bibinfo{author}{\bibfnamefont{X.-M.} \bibnamefont{Zhang}},
  \bibinfo{author}{\bibfnamefont{Z.-W.} \bibnamefont{Cui}},
  \bibinfo{author}{\bibfnamefont{X.}~\bibnamefont{Wang}}, \bibnamefont{and}
  \bibinfo{author}{\bibfnamefont{M.-H.} \bibnamefont{Yung}},
  \bibinfo{journal}{Physical Review A} \textbf{\bibinfo{volume}{97}},
  \bibinfo{pages}{052333} (\bibinfo{year}{2018}{\natexlab{a}}).

\bibitem[{\citenamefont{Yang et~al.}(2020)\citenamefont{Yang, Liu, Li, and
  Peng}}]{yang2020optimizing}
\bibinfo{author}{\bibfnamefont{X.}~\bibnamefont{Yang}},
  \bibinfo{author}{\bibfnamefont{R.}~\bibnamefont{Liu}},
  \bibinfo{author}{\bibfnamefont{J.}~\bibnamefont{Li}}, \bibnamefont{and}
  \bibinfo{author}{\bibfnamefont{X.}~\bibnamefont{Peng}},
  \bibinfo{journal}{Physical Review A} \textbf{\bibinfo{volume}{102}},
  \bibinfo{pages}{012614} (\bibinfo{year}{2020}).

\bibitem[{\citenamefont{Lin et~al.}(2020)\citenamefont{Lin, Lai, and
  Li}}]{lin2020quantum}
\bibinfo{author}{\bibfnamefont{J.}~\bibnamefont{Lin}},
  \bibinfo{author}{\bibfnamefont{Z.~Y.} \bibnamefont{Lai}}, \bibnamefont{and}
  \bibinfo{author}{\bibfnamefont{X.}~\bibnamefont{Li}},
  \bibinfo{journal}{Physical Review A} \textbf{\bibinfo{volume}{101}},
  \bibinfo{pages}{052327} (\bibinfo{year}{2020}).

\bibitem[{\citenamefont{Bukov}(2018)}]{bukov2018reinforcement}
\bibinfo{author}{\bibfnamefont{M.}~\bibnamefont{Bukov}},
  \bibinfo{journal}{Physical Review B} \textbf{\bibinfo{volume}{98}},
  \bibinfo{pages}{224305} (\bibinfo{year}{2018}).

\bibitem[{\citenamefont{Kong et~al.}(2020)\citenamefont{Kong, Zhou, Li, Yang,
  Qiu, Wu, Shi, and Du}}]{kong2020artificial}
\bibinfo{author}{\bibfnamefont{X.}~\bibnamefont{Kong}},
  \bibinfo{author}{\bibfnamefont{L.}~\bibnamefont{Zhou}},
  \bibinfo{author}{\bibfnamefont{Z.}~\bibnamefont{Li}},
  \bibinfo{author}{\bibfnamefont{Z.}~\bibnamefont{Yang}},
  \bibinfo{author}{\bibfnamefont{B.}~\bibnamefont{Qiu}},
  \bibinfo{author}{\bibfnamefont{X.}~\bibnamefont{Wu}},
  \bibinfo{author}{\bibfnamefont{F.}~\bibnamefont{Shi}}, \bibnamefont{and}
  \bibinfo{author}{\bibfnamefont{J.}~\bibnamefont{Du}}, \bibinfo{journal}{NPJ
  quantum information} \textbf{\bibinfo{volume}{6}}, \bibinfo{pages}{79}
  (\bibinfo{year}{2020}).

\bibitem[{\citenamefont{Palmieri et~al.}(2020)\citenamefont{Palmieri, Kovlakov,
  Bianchi, Yudin, Straupe, Biamonte, and Kulik}}]{palmieri2020experimental}
\bibinfo{author}{\bibfnamefont{A.~M.} \bibnamefont{Palmieri}},
  \bibinfo{author}{\bibfnamefont{E.}~\bibnamefont{Kovlakov}},
  \bibinfo{author}{\bibfnamefont{F.}~\bibnamefont{Bianchi}},
  \bibinfo{author}{\bibfnamefont{D.}~\bibnamefont{Yudin}},
  \bibinfo{author}{\bibfnamefont{S.}~\bibnamefont{Straupe}},
  \bibinfo{author}{\bibfnamefont{J.~D.} \bibnamefont{Biamonte}},
  \bibnamefont{and} \bibinfo{author}{\bibfnamefont{S.}~\bibnamefont{Kulik}},
  \bibinfo{journal}{npj Quantum Information} \textbf{\bibinfo{volume}{6}},
  \bibinfo{pages}{20} (\bibinfo{year}{2020}).

\bibitem[{\citenamefont{Wang et~al.}(2020)\citenamefont{Wang, Ashida, and
  Ueda}}]{wang2020deep}
\bibinfo{author}{\bibfnamefont{Z.~T.} \bibnamefont{Wang}},
  \bibinfo{author}{\bibfnamefont{Y.}~\bibnamefont{Ashida}}, \bibnamefont{and}
  \bibinfo{author}{\bibfnamefont{M.}~\bibnamefont{Ueda}},
  \bibinfo{journal}{Physical Review Letters} \textbf{\bibinfo{volume}{125}},
  \bibinfo{pages}{100401} (\bibinfo{year}{2020}).

\bibitem[{\citenamefont{Niu et~al.}(2019)\citenamefont{Niu, Boixo, Smelyanskiy,
  and Neven}}]{niu2019universal}
\bibinfo{author}{\bibfnamefont{M.~Y.} \bibnamefont{Niu}},
  \bibinfo{author}{\bibfnamefont{S.}~\bibnamefont{Boixo}},
  \bibinfo{author}{\bibfnamefont{V.~N.} \bibnamefont{Smelyanskiy}},
  \bibnamefont{and} \bibinfo{author}{\bibfnamefont{H.}~\bibnamefont{Neven}},
  \bibinfo{journal}{npj Quantum Information} \textbf{\bibinfo{volume}{5}},
  \bibinfo{pages}{33} (\bibinfo{year}{2019}).

\bibitem[{\citenamefont{Gratsea et~al.}(2020)\citenamefont{Gratsea, Metz, and
  Busch}}]{gratsea2020universal}
\bibinfo{author}{\bibfnamefont{A.}~\bibnamefont{Gratsea}},
  \bibinfo{author}{\bibfnamefont{F.}~\bibnamefont{Metz}}, \bibnamefont{and}
  \bibinfo{author}{\bibfnamefont{T.}~\bibnamefont{Busch}},
  \bibinfo{journal}{Journal of Physics A: Mathematical and Theoretical}
  \textbf{\bibinfo{volume}{53}}, \bibinfo{pages}{445306}
  (\bibinfo{year}{2020}).

\bibitem[{\citenamefont{Ma et~al.}(2022)\citenamefont{Ma, Dong, Ding, and
  Chen}}]{ma2022curriculum}
\bibinfo{author}{\bibfnamefont{H.}~\bibnamefont{Ma}},
  \bibinfo{author}{\bibfnamefont{D.}~\bibnamefont{Dong}},
  \bibinfo{author}{\bibfnamefont{S.~X.} \bibnamefont{Ding}}, \bibnamefont{and}
  \bibinfo{author}{\bibfnamefont{C.}~\bibnamefont{Chen}},
  \bibinfo{journal}{IEEE Transactions on Neural Networks and Learning Systems}
  (\bibinfo{year}{2022}).

\bibitem[{\citenamefont{Jordan and Mitchell}(2015)}]{jordan2015machine}
\bibinfo{author}{\bibfnamefont{M.~I.} \bibnamefont{Jordan}} \bibnamefont{and}
  \bibinfo{author}{\bibfnamefont{T.~M.} \bibnamefont{Mitchell}},
  \bibinfo{journal}{Science} \textbf{\bibinfo{volume}{349}},
  \bibinfo{pages}{255} (\bibinfo{year}{2015}).

\bibitem[{\citenamefont{Silver et~al.}(2016)\citenamefont{Silver, Huang,
  Maddison, Guez, Sifre, Van Den~Driessche, Schrittwieser, Antonoglou,
  Panneershelvam, Lanctot et~al.}}]{silver2016mastering}
\bibinfo{author}{\bibfnamefont{D.}~\bibnamefont{Silver}},
  \bibinfo{author}{\bibfnamefont{A.}~\bibnamefont{Huang}},
  \bibinfo{author}{\bibfnamefont{C.~J.} \bibnamefont{Maddison}},
  \bibinfo{author}{\bibfnamefont{A.}~\bibnamefont{Guez}},
  \bibinfo{author}{\bibfnamefont{L.}~\bibnamefont{Sifre}},
  \bibinfo{author}{\bibfnamefont{G.}~\bibnamefont{Van Den~Driessche}},
  \bibinfo{author}{\bibfnamefont{J.}~\bibnamefont{Schrittwieser}},
  \bibinfo{author}{\bibfnamefont{I.}~\bibnamefont{Antonoglou}},
  \bibinfo{author}{\bibfnamefont{V.}~\bibnamefont{Panneershelvam}},
  \bibinfo{author}{\bibfnamefont{M.}~\bibnamefont{Lanctot}},
  \bibnamefont{et~al.}, \bibinfo{journal}{nature}
  \textbf{\bibinfo{volume}{529}}, \bibinfo{pages}{484} (\bibinfo{year}{2016}).

\bibitem[{\citenamefont{Zhang et~al.}(2019{\natexlab{a}})\citenamefont{Zhang,
  Wei, Asad, Yang, and Wang}}]{zhang2019does}
\bibinfo{author}{\bibfnamefont{X.-M.} \bibnamefont{Zhang}},
  \bibinfo{author}{\bibfnamefont{Z.}~\bibnamefont{Wei}},
  \bibinfo{author}{\bibfnamefont{R.}~\bibnamefont{Asad}},
  \bibinfo{author}{\bibfnamefont{X.-C.} \bibnamefont{Yang}}, \bibnamefont{and}
  \bibinfo{author}{\bibfnamefont{X.}~\bibnamefont{Wang}}, \bibinfo{journal}{npj
  Quantum Information} \textbf{\bibinfo{volume}{5}}, \bibinfo{pages}{85}
  (\bibinfo{year}{2019}{\natexlab{a}}).

\bibitem[{\citenamefont{He et~al.}(2021{\natexlab{a}})\citenamefont{He, Wang,
  Nie, Wu, Zhang, and Wang}}]{he2021deep}
\bibinfo{author}{\bibfnamefont{R.-H.} \bibnamefont{He}},
  \bibinfo{author}{\bibfnamefont{R.}~\bibnamefont{Wang}},
  \bibinfo{author}{\bibfnamefont{S.-S.} \bibnamefont{Nie}},
  \bibinfo{author}{\bibfnamefont{J.}~\bibnamefont{Wu}},
  \bibinfo{author}{\bibfnamefont{J.-H.} \bibnamefont{Zhang}}, \bibnamefont{and}
  \bibinfo{author}{\bibfnamefont{Z.-M.} \bibnamefont{Wang}},
  \bibinfo{journal}{EPJ Quantum Technology} \textbf{\bibinfo{volume}{8}},
  \bibinfo{pages}{29} (\bibinfo{year}{2021}{\natexlab{a}}).

\bibitem[{\citenamefont{Haug et~al.}(2020)\citenamefont{Haug, Mok, You, Zhang,
  Png, and Kwek}}]{haug2020classifying}
\bibinfo{author}{\bibfnamefont{T.}~\bibnamefont{Haug}},
  \bibinfo{author}{\bibfnamefont{W.-K.} \bibnamefont{Mok}},
  \bibinfo{author}{\bibfnamefont{J.-B.} \bibnamefont{You}},
  \bibinfo{author}{\bibfnamefont{W.}~\bibnamefont{Zhang}},
  \bibinfo{author}{\bibfnamefont{C.~E.} \bibnamefont{Png}}, \bibnamefont{and}
  \bibinfo{author}{\bibfnamefont{L.-C.} \bibnamefont{Kwek}},
  \bibinfo{journal}{Machine Learning: Science and Technology}
  \textbf{\bibinfo{volume}{2}}, \bibinfo{pages}{01LT02} (\bibinfo{year}{2020}).

\bibitem[{\citenamefont{Cormen et~al.}(2022)\citenamefont{Cormen, Leiserson,
  Rivest, and Stein}}]{cormen2022introduction}
\bibinfo{author}{\bibfnamefont{T.~H.} \bibnamefont{Cormen}},
  \bibinfo{author}{\bibfnamefont{C.~E.} \bibnamefont{Leiserson}},
  \bibinfo{author}{\bibfnamefont{R.~L.} \bibnamefont{Rivest}},
  \bibnamefont{and} \bibinfo{author}{\bibfnamefont{C.}~\bibnamefont{Stein}},
  \emph{\bibinfo{title}{Introduction to algorithms}} (\bibinfo{publisher}{MIT
  press}, \bibinfo{year}{2022}).

\bibitem[{\citenamefont{Balaman}(2019)}]{balaman2019chapter}
\bibinfo{author}{\bibfnamefont{{\c{S}}.}~\bibnamefont{Balaman}},
  \bibinfo{journal}{Balaman, SYBT-D.-M.(ed.)} pp. \bibinfo{pages}{143--183}
  (\bibinfo{year}{2019}).

\bibitem[{\citenamefont{Khaneja et~al.}(2005)\citenamefont{Khaneja, Reiss,
  Kehlet, Schulte-Herbr{\"u}ggen, and Glaser}}]{khaneja2005optimal}
\bibinfo{author}{\bibfnamefont{N.}~\bibnamefont{Khaneja}},
  \bibinfo{author}{\bibfnamefont{T.}~\bibnamefont{Reiss}},
  \bibinfo{author}{\bibfnamefont{C.}~\bibnamefont{Kehlet}},
  \bibinfo{author}{\bibfnamefont{T.}~\bibnamefont{Schulte-Herbr{\"u}ggen}},
  \bibnamefont{and} \bibinfo{author}{\bibfnamefont{S.~J.}
  \bibnamefont{Glaser}}, \bibinfo{journal}{Journal of magnetic resonance}
  \textbf{\bibinfo{volume}{172}}, \bibinfo{pages}{296} (\bibinfo{year}{2005}).

\bibitem[{\citenamefont{Rowland and Jones}(2012)}]{rowland2012implementing}
\bibinfo{author}{\bibfnamefont{B.}~\bibnamefont{Rowland}} \bibnamefont{and}
  \bibinfo{author}{\bibfnamefont{J.~A.} \bibnamefont{Jones}},
  \bibinfo{journal}{Philosophical Transactions of the Royal Society A:
  Mathematical, Physical and Engineering Sciences}
  \textbf{\bibinfo{volume}{370}}, \bibinfo{pages}{4636} (\bibinfo{year}{2012}).

\bibitem[{\citenamefont{Doria et~al.}(2011)\citenamefont{Doria, Calarco, and
  Montangero}}]{doria2011optimal}
\bibinfo{author}{\bibfnamefont{P.}~\bibnamefont{Doria}},
  \bibinfo{author}{\bibfnamefont{T.}~\bibnamefont{Calarco}}, \bibnamefont{and}
  \bibinfo{author}{\bibfnamefont{S.}~\bibnamefont{Montangero}},
  \bibinfo{journal}{Physical review letters} \textbf{\bibinfo{volume}{106}},
  \bibinfo{pages}{190501} (\bibinfo{year}{2011}).

\bibitem[{\citenamefont{Caneva et~al.}(2011)\citenamefont{Caneva, Calarco, and
  Montangero}}]{caneva2011chopped}
\bibinfo{author}{\bibfnamefont{T.}~\bibnamefont{Caneva}},
  \bibinfo{author}{\bibfnamefont{T.}~\bibnamefont{Calarco}}, \bibnamefont{and}
  \bibinfo{author}{\bibfnamefont{S.}~\bibnamefont{Montangero}},
  \bibinfo{journal}{Physical Review A} \textbf{\bibinfo{volume}{84}},
  \bibinfo{pages}{022326} (\bibinfo{year}{2011}).

\bibitem[{\citenamefont{He et~al.}(2021{\natexlab{b}})\citenamefont{He, Liu,
  Wang, Wu, Nie, and Wang}}]{he2021universal}
\bibinfo{author}{\bibfnamefont{R.-H.} \bibnamefont{He}},
  \bibinfo{author}{\bibfnamefont{H.-D.} \bibnamefont{Liu}},
  \bibinfo{author}{\bibfnamefont{S.-B.} \bibnamefont{Wang}},
  \bibinfo{author}{\bibfnamefont{J.}~\bibnamefont{Wu}},
  \bibinfo{author}{\bibfnamefont{S.-S.} \bibnamefont{Nie}}, \bibnamefont{and}
  \bibinfo{author}{\bibfnamefont{Z.-M.} \bibnamefont{Wang}},
  \bibinfo{journal}{Quantum Science and Technology}
  \textbf{\bibinfo{volume}{6}}, \bibinfo{pages}{045021}
  (\bibinfo{year}{2021}{\natexlab{b}}).

\bibitem[{\citenamefont{Zhang et~al.}(2018{\natexlab{b}})\citenamefont{Zhang,
  Li, Wang, Cao, Xiao, and Guo}}]{zhang2018qubits}
\bibinfo{author}{\bibfnamefont{X.}~\bibnamefont{Zhang}},
  \bibinfo{author}{\bibfnamefont{H.-O.} \bibnamefont{Li}},
  \bibinfo{author}{\bibfnamefont{K.}~\bibnamefont{Wang}},
  \bibinfo{author}{\bibfnamefont{G.}~\bibnamefont{Cao}},
  \bibinfo{author}{\bibfnamefont{M.}~\bibnamefont{Xiao}}, \bibnamefont{and}
  \bibinfo{author}{\bibfnamefont{G.-P.} \bibnamefont{Guo}},
  \bibinfo{journal}{Chinese Physics B} \textbf{\bibinfo{volume}{27}},
  \bibinfo{pages}{020305} (\bibinfo{year}{2018}{\natexlab{b}}).

\bibitem[{\citenamefont{Petta et~al.}(2005)\citenamefont{Petta, Johnson,
  Taylor, Laird, Yacoby, Lukin, Marcus, Hanson, and
  Gossard}}]{petta2005coherent}
\bibinfo{author}{\bibfnamefont{J.~R.} \bibnamefont{Petta}},
  \bibinfo{author}{\bibfnamefont{A.~C.} \bibnamefont{Johnson}},
  \bibinfo{author}{\bibfnamefont{J.~M.} \bibnamefont{Taylor}},
  \bibinfo{author}{\bibfnamefont{E.~A.} \bibnamefont{Laird}},
  \bibinfo{author}{\bibfnamefont{A.}~\bibnamefont{Yacoby}},
  \bibinfo{author}{\bibfnamefont{M.~D.} \bibnamefont{Lukin}},
  \bibinfo{author}{\bibfnamefont{C.~M.} \bibnamefont{Marcus}},
  \bibinfo{author}{\bibfnamefont{M.~P.} \bibnamefont{Hanson}},
  \bibnamefont{and} \bibinfo{author}{\bibfnamefont{A.~C.}
  \bibnamefont{Gossard}}, \bibinfo{journal}{Science}
  \textbf{\bibinfo{volume}{309}}, \bibinfo{pages}{2180} (\bibinfo{year}{2005}).

\bibitem[{\citenamefont{Levy}(2002)}]{levy2002universal}
\bibinfo{author}{\bibfnamefont{J.}~\bibnamefont{Levy}},
  \bibinfo{journal}{Physical Review Letters} \textbf{\bibinfo{volume}{89}},
  \bibinfo{pages}{147902} (\bibinfo{year}{2002}).

\bibitem[{\citenamefont{Malinowski et~al.}(2017)\citenamefont{Malinowski,
  Martins, Nissen, Barnes, Cywi{\'n}ski, Rudner, Fallahi, Gardner, Manfra,
  Marcus et~al.}}]{malinowski2017notch}
\bibinfo{author}{\bibfnamefont{F.~K.} \bibnamefont{Malinowski}},
  \bibinfo{author}{\bibfnamefont{F.}~\bibnamefont{Martins}},
  \bibinfo{author}{\bibfnamefont{P.~D.} \bibnamefont{Nissen}},
  \bibinfo{author}{\bibfnamefont{E.}~\bibnamefont{Barnes}},
  \bibinfo{author}{\bibfnamefont{{\L}.}~\bibnamefont{Cywi{\'n}ski}},
  \bibinfo{author}{\bibfnamefont{M.~S.} \bibnamefont{Rudner}},
  \bibinfo{author}{\bibfnamefont{S.}~\bibnamefont{Fallahi}},
  \bibinfo{author}{\bibfnamefont{G.~C.} \bibnamefont{Gardner}},
  \bibinfo{author}{\bibfnamefont{M.~J.} \bibnamefont{Manfra}},
  \bibinfo{author}{\bibfnamefont{C.~M.} \bibnamefont{Marcus}},
  \bibnamefont{et~al.}, \bibinfo{journal}{Nature nanotechnology}
  \textbf{\bibinfo{volume}{12}}, \bibinfo{pages}{16} (\bibinfo{year}{2017}).

\bibitem[{\citenamefont{Foletti et~al.}(2009)\citenamefont{Foletti, Bluhm,
  Mahalu, Umansky, and Yacoby}}]{foletti2009universal}
\bibinfo{author}{\bibfnamefont{S.}~\bibnamefont{Foletti}},
  \bibinfo{author}{\bibfnamefont{H.}~\bibnamefont{Bluhm}},
  \bibinfo{author}{\bibfnamefont{D.}~\bibnamefont{Mahalu}},
  \bibinfo{author}{\bibfnamefont{V.}~\bibnamefont{Umansky}}, \bibnamefont{and}
  \bibinfo{author}{\bibfnamefont{A.}~\bibnamefont{Yacoby}},
  \bibinfo{journal}{Nature Physics} \textbf{\bibinfo{volume}{5}},
  \bibinfo{pages}{903} (\bibinfo{year}{2009}).

\bibitem[{\citenamefont{Zhang et~al.}(2019{\natexlab{b}})\citenamefont{Zhang,
  Li, Cao, Xiao, Guo, and Guo}}]{zhang2019semiconductor}
\bibinfo{author}{\bibfnamefont{X.}~\bibnamefont{Zhang}},
  \bibinfo{author}{\bibfnamefont{H.-O.} \bibnamefont{Li}},
  \bibinfo{author}{\bibfnamefont{G.}~\bibnamefont{Cao}},
  \bibinfo{author}{\bibfnamefont{M.}~\bibnamefont{Xiao}},
  \bibinfo{author}{\bibfnamefont{G.-C.} \bibnamefont{Guo}}, \bibnamefont{and}
  \bibinfo{author}{\bibfnamefont{G.-P.} \bibnamefont{Guo}},
  \bibinfo{journal}{National Science Review} \textbf{\bibinfo{volume}{6}},
  \bibinfo{pages}{32} (\bibinfo{year}{2019}{\natexlab{b}}).

\bibitem[{\citenamefont{Shulman et~al.}(2012)\citenamefont{Shulman, Dial,
  Harvey, Bluhm, Umansky, and Yacoby}}]{shulman2012demonstration}
\bibinfo{author}{\bibfnamefont{M.~D.} \bibnamefont{Shulman}},
  \bibinfo{author}{\bibfnamefont{O.~E.} \bibnamefont{Dial}},
  \bibinfo{author}{\bibfnamefont{S.~P.} \bibnamefont{Harvey}},
  \bibinfo{author}{\bibfnamefont{H.}~\bibnamefont{Bluhm}},
  \bibinfo{author}{\bibfnamefont{V.}~\bibnamefont{Umansky}}, \bibnamefont{and}
  \bibinfo{author}{\bibfnamefont{A.}~\bibnamefont{Yacoby}},
  \bibinfo{journal}{science} \textbf{\bibinfo{volume}{336}},
  \bibinfo{pages}{202} (\bibinfo{year}{2012}).

\bibitem[{\citenamefont{Wang et~al.}(2015)\citenamefont{Wang, Barnes, and
  Sarma}}]{wang2015improving}
\bibinfo{author}{\bibfnamefont{X.}~\bibnamefont{Wang}},
  \bibinfo{author}{\bibfnamefont{E.}~\bibnamefont{Barnes}}, \bibnamefont{and}
  \bibinfo{author}{\bibfnamefont{S.~D.} \bibnamefont{Sarma}},
  \bibinfo{journal}{npj Quantum Information} \textbf{\bibinfo{volume}{1}},
  \bibinfo{pages}{1} (\bibinfo{year}{2015}).

\bibitem[{\citenamefont{Van~Weperen et~al.}(2011)\citenamefont{Van~Weperen,
  Armstrong, Laird, Medford, Marcus, Hanson, and Gossard}}]{van2011charge}
\bibinfo{author}{\bibfnamefont{I.}~\bibnamefont{Van~Weperen}},
  \bibinfo{author}{\bibfnamefont{B.}~\bibnamefont{Armstrong}},
  \bibinfo{author}{\bibfnamefont{E.}~\bibnamefont{Laird}},
  \bibinfo{author}{\bibfnamefont{J.}~\bibnamefont{Medford}},
  \bibinfo{author}{\bibfnamefont{C.}~\bibnamefont{Marcus}},
  \bibinfo{author}{\bibfnamefont{M.}~\bibnamefont{Hanson}}, \bibnamefont{and}
  \bibinfo{author}{\bibfnamefont{A.}~\bibnamefont{Gossard}},
  \bibinfo{journal}{Physical review letters} \textbf{\bibinfo{volume}{107}},
  \bibinfo{pages}{030506} (\bibinfo{year}{2011}).

\bibitem[{\citenamefont{Xie et~al.}(2022)\citenamefont{Xie, Ren, He, Ablimit,
  and Wang}}]{yangyang}
\bibinfo{author}{\bibfnamefont{Y.-Y.} \bibnamefont{Xie}},
  \bibinfo{author}{\bibfnamefont{F.-H.} \bibnamefont{Ren}},
  \bibinfo{author}{\bibfnamefont{R.-H.} \bibnamefont{He}},
  \bibinfo{author}{\bibfnamefont{A.}~\bibnamefont{Ablimit}}, \bibnamefont{and}
  \bibinfo{author}{\bibfnamefont{Z.-M.} \bibnamefont{Wang}},
  \bibinfo{journal}{Phys. Rev. A} \textbf{\bibinfo{volume}{106}},
  \bibinfo{pages}{062612} (\bibinfo{year}{2022}).

\end{thebibliography}

\end{document}